\documentclass[twocolumn, trackchanges]{aastex701}

\usepackage{amsmath}
\usepackage{graphicx}
\graphicspath{{figures/}}

\newcommand{\revise}[1]{\textcolor{black}{#1}}

\begin{document}

\title{Free Energy Sources of Ion-scale Waves Observed by Parker Solar Probe}

\author[orcid=0000-0002-8941-3463]{Niranjana Shankarappa}
\affiliation{Lunar and Planetary Laboratory, University of Arizona, Tucson, AZ 85721, USA}
\email{niranjanats@arizona.edu}  

\author[orcid=0000-0001-6038-1923]{Kristopher G. Klein}
\affiliation{Lunar and Planetary Laboratory, University of Arizona, Tucson, AZ 85721, USA}
\email{kgklein@arizona.edu}

\author[orcid=0000-0002-7365-0472]{Mihailo M. Martinovi\'c}
\affiliation{Lunar and Planetary Laboratory, University of Arizona, Tucson, AZ 85721, USA}
\email{mmartinovic@arizona.edu}

\author[orcid=0000-0002-4625-3332]{Trevor A. Bowen}
\affiliation{Space Sciences Laboratory, University of California, Berkeley, CA 94720, USA}
\email{tbowen@berkeley.edu}

\author[orcid=0000-0001-5030-6030]{Davin E. Larson}
\affiliation{Space Sciences Laboratory, University of California, Berkeley, CA 94720, USA}
\email{davin@ssl.berkeley.edu}

\author[orcid=0000-0002-0396-0547]{Roberto Livi}
\affiliation{Space Sciences Laboratory, University of California, Berkeley, CA 94720, USA}
\email{robertolivi@gmail.com}

\author[orcid=0000-0003-0519-6498]{Ali Rahmati}
\affiliation{Space Sciences Laboratory, University of California, Berkeley, CA 94720, USA}
\email{rahmati@berkeley.edu}

\author[orcid=0000-0002-7287-5098]{Phyllis L. Whittlesey}
\affiliation{Space Sciences Laboratory, University of California, Berkeley, CA 94720, USA}
\email{phyllisw@berkeley.edu}

\author[orcid=0000-0002-7728-0085]{Michael L. Stevens}
\affiliation{Smithsonian Astrophysical Observatory, Cambridge, MA 02138, USA}
\email{mstevens@cfa.harvard.edu}

\begin{abstract}

Parker Solar Probe (PSP) observes abundant circularly polarized ion-scale waves throughout the inner heliosphere. 
These waves are a signature of the interplay between plasma microinstabilities and turbulent dissipation. 
We perform a mission-wide statistical survey of ion-scale waves observed by PSP, investigating if the waves correspond to specific free energy sources in the measured proton velocity distributions. 
We find that left-handed waves (LHWs) are frequently observed, with the fraction of time they are observed increasing closer to the Sun, reaching $\sim$30\%. 
Right-handed waves (RHWs) are less frequently observed, with the associated time fraction decreasing closer to the Sun. 
The observed LHWs are generally consistent with parallel propagating ion cyclotron wave (ICW) storms that occur continuously for extended periods of time. 
Turbulent energy spectra are consistently steeper when LHW storms are observed; these wave storms mediate the spatial transport of the free energy associated with temperature anisotropy. 
The observed RHWs are generally consistent with oblique and parallel fast magnetosonic waves (FMWs), and their observation is well correlated with enhanced proton parallel heat flux, which quantifies the presence of secondary proton populations. 
Using observations and the \texttt{SAVIC} machine learning instability identification algorithm, we identify a threshold on the proton heat flux beyond which FMWs are likely to be driven unstable by the proton beams. 
We are thus able to associate trends in the observed ion-scale waves with known sources of free energy for Encounters 3 through 24 of the PSP's prime science phase.

\end{abstract}


\section{Introduction} 

Understanding what processes mediate the extended heating of the solar wind is a long-standing problem in space plasma physics. 
A combination of dissipation of turbulence \citep{Matthaeus_1999, Howes_2024} and ion-scale waves \citep{Leamon_1998_ICW, Hollweg_Isenberg_2002_wave_heating,Bowen_2022_Cyclotron_dissipation, Bowen_2024_extended_cylotron_heating, Niranjana_2024} likely contribute to this extended heating \citep{Cranmer_2003, Cranmer_2007, Cranmer_2014}. 
Turbulence is hypothesized to be initiated by outward-propagating large-scale Alfv\'en waves \citep{Belcher_1969_Alfven_waves} which cascade wavevector-anisotropically \citep{Matthaeus_1990_k_aniso, GS_1995_critical_balance, Horbury2012_k_aniso, Chen_2016_Review} to smaller scales, transitioning to a kinetic-Alfv\'en wave cascade \citep{Leamon_1998_KAWs, Salem_2012_KAW_turb, Schekochihin_2022_biased_review} at kinetic scales where turbulent energy can be dissipated onto ions and electrons via wave particle resonant mechanisms like Landau damping \citep{Leamon_1999, Chen_2019_LD} and non-resonant mechanisms like stochastic heating \citep{Chen_2001_SH, Chandran_2010_stochastic_heating}. 
The imbalanced turbulent cascade could also be channeled to ion cyclotron waves (ICWs) through the “helicity barrier” mechanism \citep{Meyrand_2021,Squire_2022_helicity_barrier, Mcintyre_2025_helicity_barrier}, where the driven ICWs undergo cyclotron damping onto ions. 
Furthermore, turbulence can generate intermittent structures \citep{Dmitruk_2004, Matthaeus_Velli_2011} at kinetic scales, which can dissipate through kinetic processes such as magnetic reconnection \citep{Vech_2018_MRX}. 

Non-Maxwellian ion velocity distribution functions (VDFs) are frequently observed in the solar wind (\cite{Marsch_2012_pVDF_Helios}, see the review in \cite{Verscharen_2019_Review}). 
Such distributions have sources of free energy, including anisotropies in temperatures along parallel and perpendicular directions with respect to the local magnetic field and the presence of drifting secondary populations. 
The observed deviation from local thermal equilibrium (LTE) is due to a combination of processes: anisotropic expansion of solar wind \citep{CGL_SW_expansion}; preferential heating of ions in the direction perpendicular to the local magnetic field at closer distances to the Sun \citep{Maruca_2011_T_p_aniso, Kasper_and_Klein_2019} possibly due to mechanisms like stochastic heating, cyclotron damping, and the helicity barrier. 
Moreover, preferential heating of alpha particles \citep{Robbins_1970_alpha_differential_streaming, Mostafavi_2022_alpha_diff_streaming} and proton beams \citep{Verniero_2020_waves_proton_beams, Phan_2022_p_beams_reconnection} can drive a differential flow with respect to core protons \citep{Alterman_2018_proton_beams_alphas_WIND}. 
The non-Maxwellian VDFs are constrained by microinstabilities that generate ion-scale circularly polarized waves, i.e., ICWs and fast magnetosonic waves (FMWs) as well as oblique non-circularly polarized waves (see review in \cite{Verscharen_2019_Review}). 
These ion-scale waves mediate the local transport of non-LTE free energy and may heat the solar wind. 

Parker Solar Probe (PSP, \cite{Fox:2015}), launched in 2018, has completed 24 orbital encounters as of June 19, 2025, observing the Sun at distances as close as $9.86 R_\odot$ (solar radii).  
A remarkable observation made by PSP is that ion-scale circularly polarized waves are abundant in the inner heliosphere \citep{Bowen_2020_ICW, Liu_2023, Niranjana_2024}. 
The observed ion-scale waves are likely the result of a combination of above-mentioned preferential dissipation processes and instabilities associated with non-LTE ion VDFs. 

In this work, we investigate whether known sources of free energy are consistent with observed waves. 
We use in situ PSP observations of solar wind thermal plasma \citep{Kasper2016} and electromagnetic fields \citep{Bale2016} (see Sections \ref{sec:meth_PSP_FIELDS} and \ref{sec:meth_PSP_SPAN}) to study the statistical trends of both ion-scale waves and non-Maxwellian features of proton VDFs from all available PSP Encounters. 
We find that, consistent with \cite{Bowen_2020_ICW} and \cite{Liu_2023}, left-handed waves (LHWs) are dominantly observed and right-handed waves (RHWs) are less frequently seen. 
LHW storms occur continuously over long time periods, up to several days. 
Whether the LHWs are observed or not observed is well correlated with the sampling angle between the local mean magnetic field and the solar wind velocity being smaller or larger. This dependence of wave observation on sampling angle is consistent with previous studies that have analyzed solar wind observations from various spacecraft \citep{Murphy_1995_wave_sampling_angle, Jian_2009, Jian_2014, Boardsen_2015, Bowen_2020_ICW} 
%Here, the sampling angle is the  in the spacecraft frame. 
The LHWs are generally consistent with anti-sunward, parallel propagating ICWs (in agreement with \cite{Bowen_2020_outward_ICWs}). 
The RHWs are seen in bursty patches, with their observation well-correlated with the presence of proton beams and weakly correlated with sampling angle, making them consistent with parallel and oblique propagating FMWs. 
%Furthermore, the observed LHW storms are well correlated with steeper turbulent energy spectra at dissipation scales, which indicates that they mediate processes like cyclotron damping, stochastic heating, and helicity barrier. 
%The observed RHWs are coincident with beams that are possibly generated by reconnection near the heliospheric current sheet. 
Using the observed RHW dependence on the proton parallel heat flux and the \texttt{SAVIC} machine learning plasma instability identification algorithm (\cite{Martinović_2023_SAVIC}, see Appendix \ref{sec:meth_SAVIC}), we infer a threshold on the proton parallel heat flux above which the proton beams are likely to drive unstable FMWs. 
This work provides observational constraints to theories that describe the processes that are mediated by ion-scale waves (e.g., \cite{Chandran_2010_oblique_par_ICWs}, \cite{Squire_2022_helicity_barrier}, \cite{Bowen_2024_waves_intermittency}, \cite{Yerger_2024_AGU}).

\section{Methodology} 
This section provides a brief overview of the procedure employed in this work. 
The methodology is described in more detail in Appendix \ref{sec:methodology}.
We use PSP/FIELDS from Encounters 1-24 and PSP/SPANi observations from Encounters 3-24, when available. 
Encounters encompass PSP observations during the $\sim$11 days around perihelion when the sampling rates of FIELDS MAG and SPANi are maximal. 
The PSP observations during Encounters 1-24 cover radial distances of 0.05-0.3 au.

We use PSP/FIELDS \citep{Bale2016} magnetic field (\textbf{B}) observations and divide them into continuous 15-minute intervals. 
We downsample \textbf{B} time series to the frequencies at which observed \textbf{B} spectra hit the MAG noise floor \citep{Bowen_noise} and then evaluate $\sigma(s,t)$, the spectrum of spacecraft frame circular polarization of magnetic fluctuations using a wavelet transform as a function of inverse frequency, $s$, and time $t$. 
The $\sigma(s,t)$ values are downsampled and stored at a cadence of a New York second (1 NYs$\sim 0.874$ s). 
We then distinguish the frequency-time regions corresponding to ion-scale circularly polarized waves from background turbulence using $|\sigma(s,t)| = 0.7$ as a threshold, allowing separate evaluation of the corresponding ion-scale wave and turbulent energy spectra. 
Upon repeating the wave identification routine for 15-minute intervals from all observations from Encounters 1- 24, we produce a mission-wide ion-scale wave repository \citep{Niranjana_2025_Zenodo}, which documents the presence or absence of these waves and the frequency band at which they are observed at a cadence of a NYs. 
Using the wave repository, the time fraction of wave observation, $F_{wave}$, is evaluated at multiple cadences as required by different analysis methods. 
Additionally, we estimate $\alpha_{max}$, the steepest slope in the dissipation region of the evaluated turbulent spectra. 
Note that the wave energy spectra has been removed to evaluate $\alpha_{max}$ (similar to \cite{Mcintyre_2025_helicity_barrier}). 
Here the dissipation region is estimated as the frequency band between 0.1 Hz and the frequency at which MAG hits the noise floor.
The processing of FIELDS data to create the wave repository is described in more detail in Appendix \ref{sec:meth_PSP_FIELDS}. 

We then use PSP/SPANi \citep{Livi_2022} observations to evaluate non-Maxwellian features of proton VDFs, $f_p$, specifically the temperature anisotropy ($T_{\perp, p}/T_{\parallel,p}$) and normalized parallel heat flux ($q_{\parallel, p}/q_{0,p}$) as described in Appendix \ref{sec:meth_PSP_SPAN}. 
The $T_{\perp, p}/T_{\parallel,p}$ and $q_{\parallel, p}/q_{0,p}$ values are discarded if the SPANi goodness of field-of-view (FOV) criterion (described in Appendix \ref{sec:meth_SPAN_FOV}) is not satisfied. 
Moreover, QTN electron densities ($n_{e,QTN}$) are preferred over SPANi proton densities ($n_p$) to evaluate parameters that depend on proton density, assuming that $n_{e,QTN} \simeq n_p$ and neglecting the contribution of alpha particles to the density. 
We then analyze the dependence of wave occurrence on the proton temperature anisotropy instabilities as described in Appendix \ref{sec:meth_Tp_aniso_inst} and Section \ref{sec:F_wave_Vs_T_p_aniso}. 
We further employ the \texttt{SAVIC} machine learning algorithm to analyze the dependence of wave observations on proton beam instabilities, as described in Appendix \ref{sec:meth_SAVIC} and Section \ref{sec:F_wave_Vs_q_par}.

\section{Encounter-wise trends of ion-scale waves and solar wind parameters} \label{sec:encounterwise-trend}

\begin{figure*}[hbt!]
    \centering
    \includegraphics[width = 0.95\linewidth]{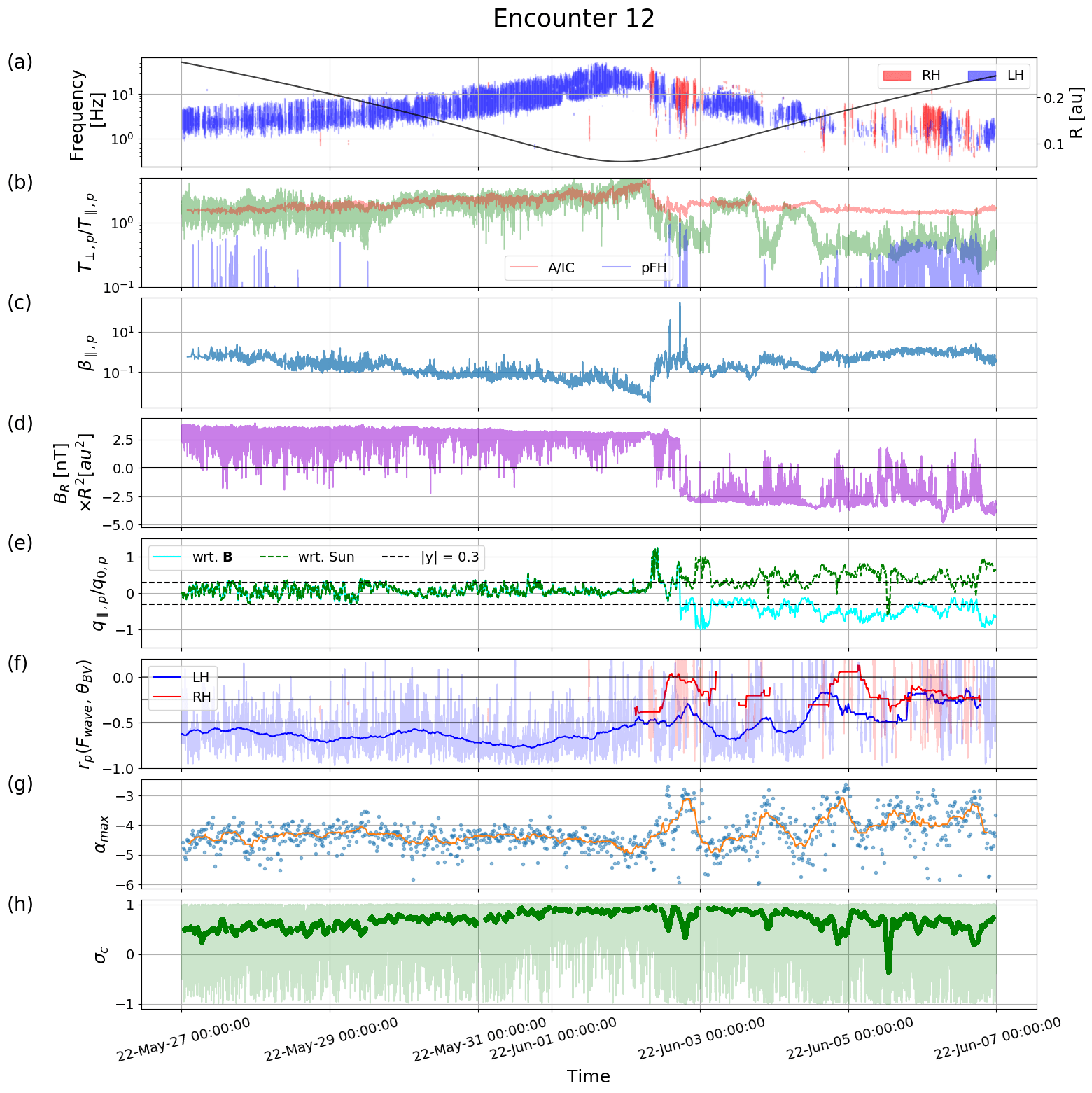}
    \caption{Encounter-wise trends of observed ion-scale waves and solar wind parameters for Encounter 12: (a) frequency of observed ion-scale LHWs (blue) and RHWs (red); (b) observed proton temperature anisotropy ($T_{\perp,p}/T_{\parallel,p}$, green) evaluated from SPANi proton VDFs along with ion cyclotron instability threshold (red) and parallel firehose instability threshold (blue); (c) the proton parallel plasma beta ($\beta_{\parallel,p}$, blue); (d) the normalized radial component of magnetic field ($B_R$, purple); (e) moving median of normalized proton parallel heat flux ($q_{\parallel,p}/q_{0,p}$) evaluated using SPANi proton VDFs where the cyan (green) curve is $q_{\parallel,p}/q_{0,p}$ with respect to local \textbf{B} direction (the Sun); (f) the Pearson correlation coefficient between sampling angle ($\theta_{BV}$) and time fraction of observation ($F_{wave}$) of LHWs (blue shaded) and RHWs (red shaded) along with their moving medians (solid curves); (g) the steepest slope in the transition region ($\alpha_{max}$, blue) of observed turbulent energy spectra and it’s moving median (orange); and (h) the cross helicity ($\sigma_c$) along with it’s moving mean (green).
}
    \label{fig:E12_encounterwise_trends}
\end{figure*}

In this subsection, we present the ion-scale wave observations as a function of time for a representative encounter and study the dependence of waves on solar wind and plasma parameters. 
Figure \ref{fig:E12_encounterwise_trends} presents an overview of Encounter 12. 
Ion-scale wave observations of the encounter are divided into 5-minute intervals, and the frequency ranges at which LHWs and RHWs are observed within each 5-minute interval are evaluated.
Panel (a) illustrates the identified LHW (blue) and RHW (red) frequencies as a function of the median time of all 5-minute intervals in the encounter. 
This color scheme for LHWs/RHWs is used throughout this paper.
Note that the duration of the waves identified within each 5 minute interval can be much shorter than 5 minutes. 
Hence, panel (a) does not necessarily represent how frequently waves are observed, and the latter is represented by the $F_{wave}$ parameter (see panel (b) of Figure \ref{fig:F_wave_Vs_R_and_wave_energy_Vs_R} and Section \ref{sec:F_wave_R_profile}). 
In panel (b), the proton temperature anisotropy ($T_{\perp,p}/T_{\parallel,p}$, green) evaluated from the observed SPANi proton VDFs (as described in Appendix \ref{sec:meth_PSP_SPAN}) is plotted as a function of time along with thresholds of ion cyclotron (red) and parallel firehose (blue) instabilities (see Appendix \ref{sec:meth_Tp_aniso_inst}). 
The proton parallel plasma beta, $\beta_{\parallel,p}$, is shown in panel (c), and the normalized radial component of the magnetic field averaged over each SPANi measurement is displayed in panel (d). 
Panel (e) illustrates the moving mean of the evaluated normalized proton parallel heat flux, $q_{\parallel,p}/q_{0,p}$ (see Appendix \ref{sec:meth_PSP_SPAN}). 
Here, the green curve is $q_{\parallel,p}/q_{0,p}$ with respect to the Sun, and the cyan line is $q_{\parallel,p}/q_{0,p}$ renormalized with respect to the local \textbf{B} direction. 
A positive value of the cyan (green) curve represents heat flux along the local \textbf{B} (anti-sunwards), and a negative value represents heat flux opposite to local \textbf{B} (sunwards). 
The black dashed lines correspond to the threshold value of $|q_{\parallel,p}/q_{0,p}| = 0.3$ above which FMWs are likely to be driven unstable by proton beams (see Section \ref{sec:F_wave_Vs_q_par}).
In panel (f), the shaded region is the Pearson correlation coefficient between $F_{wave}$ and the sampling angle ($\theta_{BV}$) for LHW (blue) and RHW (red) and the solid lines are the moving means of the shaded regions (see Section \ref{sec:_meth_wave_theta_anti_corr} for a discussion on $F_{wave}-\theta_{BV}$ correlation). 
Panel (g) shows the steepest slope in the kinetic scales of observed turbulent energy spectra ($\alpha_{max}$, blue; see Section \ref{sec:meth_PSP_FIELDS}) of continuous 15-minute intervals as a function of time along with the moving median (orange). 
Panel (h) displays the cross helicity values (green) evaluated at a frequency of 0.01 Hz using Equations 3-10 from \cite{Wicks_2013} and SPANi $n_p$. 
Note that using $n_{e,QTN}$ to evaluate cross-helicity instead of SPANi $n_p$ produces qualitatively similar radial profiles.
The goodness of FOV criteria have not been applied to the SPANi observations in the encounter-wise trend plots, as such a selection would make it difficult to infer temporal trends by excluding large chunks of observations, but are enforced in the statistical analysis starting in Section~\ref{sec:F_wave_R_profile}.

From Figure \ref{fig:E12_encounterwise_trends}, we make the following observations. 
LHW storms occur over extended periods of time, at times up to several days, as seen in panel (a). The observation of LHWs dominantly depends on the sampling angle, as illustrated by the consistently large negative Pearson correlation coefficient between $F^{LH}_{wave}$ and $\theta_{BV}$ (blue curve panel (f)).
During such LHW storms, the absence of LHWs is most likely due to the sampling angle being large. Therefore, we argue that the LHWs are parallel-propagating and occur continuously over a wide range of radial distances and Carrington latitudes. They are signatures of processes that constantly generate and damp LHWs and these processes are active in solar wind emanating from different source regions. 
However, the LHW observation also depends on the radial distance (see Figure \ref{fig:F_wave_Vs_R_and_wave_energy_Vs_R}). 
 The LHWs are more likely to occur invariably at closer radial distances. 
Moreover, the observed temperature anisotropy is consistently close to the ion cyclotron instability threshold throughout the LHW storms as seen in panel (b), indicating that the LHW storms are ICWs. Because these ICWs retain their left-handed plasma frame polarization, they are anti-sunward propagating \citep{Bowen_2020_outward_ICWs, Niranjana_2024}.
Furthermore, the slope of turbulent energy spectra, $\alpha_{max}$, is consistently steeper during LHW storms as has been demonstrated previously in \cite{Bowen_2024_extended_cylotron_heating} and \cite{Mcintyre_2025_helicity_barrier} on more limited data sets. 
The incessant parallel propagating ICW wave storms are only interrupted by relatively transient bursts of RHWs. While the observation of RHWs is weakly correlated with the sampling angle (red curve, panel (f)), it is well correlated with enhancements in proton parallel heat flux as seen in panel (e). 
The normalized heat flux strikingly increases when RHWs are observed and remains consistently low during LHW storms. 
A higher value of normalized heat flux represents the presence of proton beams, and thus the observed RHWs are possibly parallel and oblique FMWs that are generated due to proton beam FMW instability. 
Moreover, $\alpha_{max}$ is shallow when RHWs are observed. 
Furthermore, the value of cross helicity is lower when RHWs are observed, indicating the presence of a non-Alfv\'enic structure (possibly the heliocentric current sheet (HCS)). 
These observed trends are generally consistent for observations from Encounter 10 onward. 
The encounter-wise trends for the rest of Encounters 3-24 are presented in the Appendix~\ref{sec:Ewise}.

\section{Radial profiles of Ion-scale Wave duration and energy} \label{sec:F_wave_R_profile}

\begin{figure}[hbt!]
    \centering
    \includegraphics[width=0.95\linewidth]{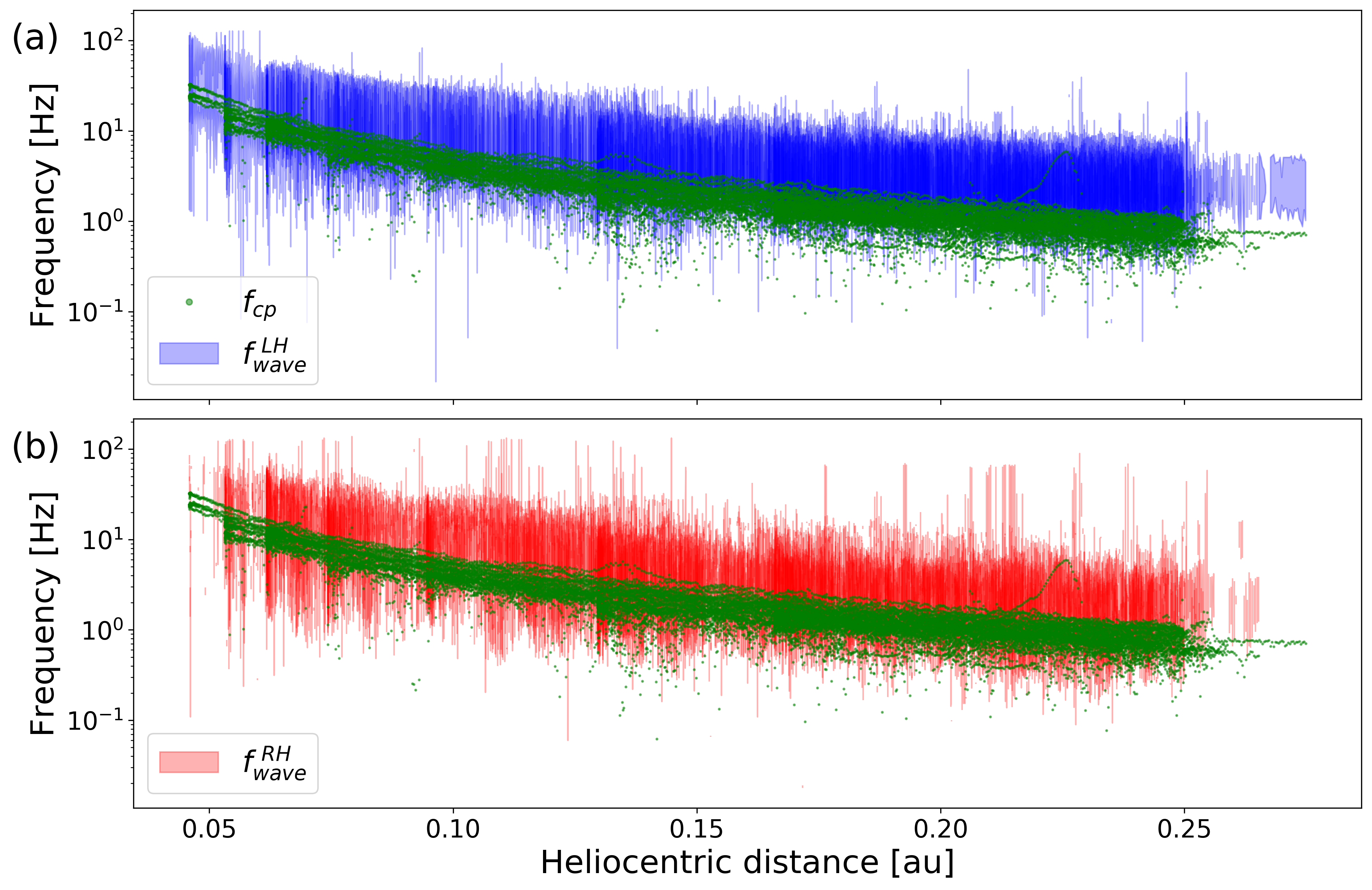}
    \caption{Radial profile of ion scale LHW (a) and RHW (b) frequencies for PSP observations from Encounter 1-24. 
    The local proton cyclotron frequencies are plotted in green.}
    \label{fig:f_wave_LH_RH_fcp}
\end{figure}

Using the wave repository \citep{Niranjana_2025_Zenodo} where the frequency band of LHWs and RHWs ($f_{wave}$) is recorded as a function of time, the radial variation of the ion-scale LHW (blue) and RHW (red) frequencies for Encounters 1-24 is shown in Figure \ref{fig:f_wave_LH_RH_fcp}.  
The frequencies of observed LHWs and RHWs are comparable to local proton cyclotron frequencies (green) as expected for Doppler-shifted ion-scale waves \citep{Jian_2009, Bowen_2020_ICW, Niranjana_2024}. 
All of the red, blue, and green curves scale with decreasing magnetic field magnitude with increasing heliocentric distance. 

\begin{figure*}
    \centering
    \includegraphics[width=0.75\linewidth]{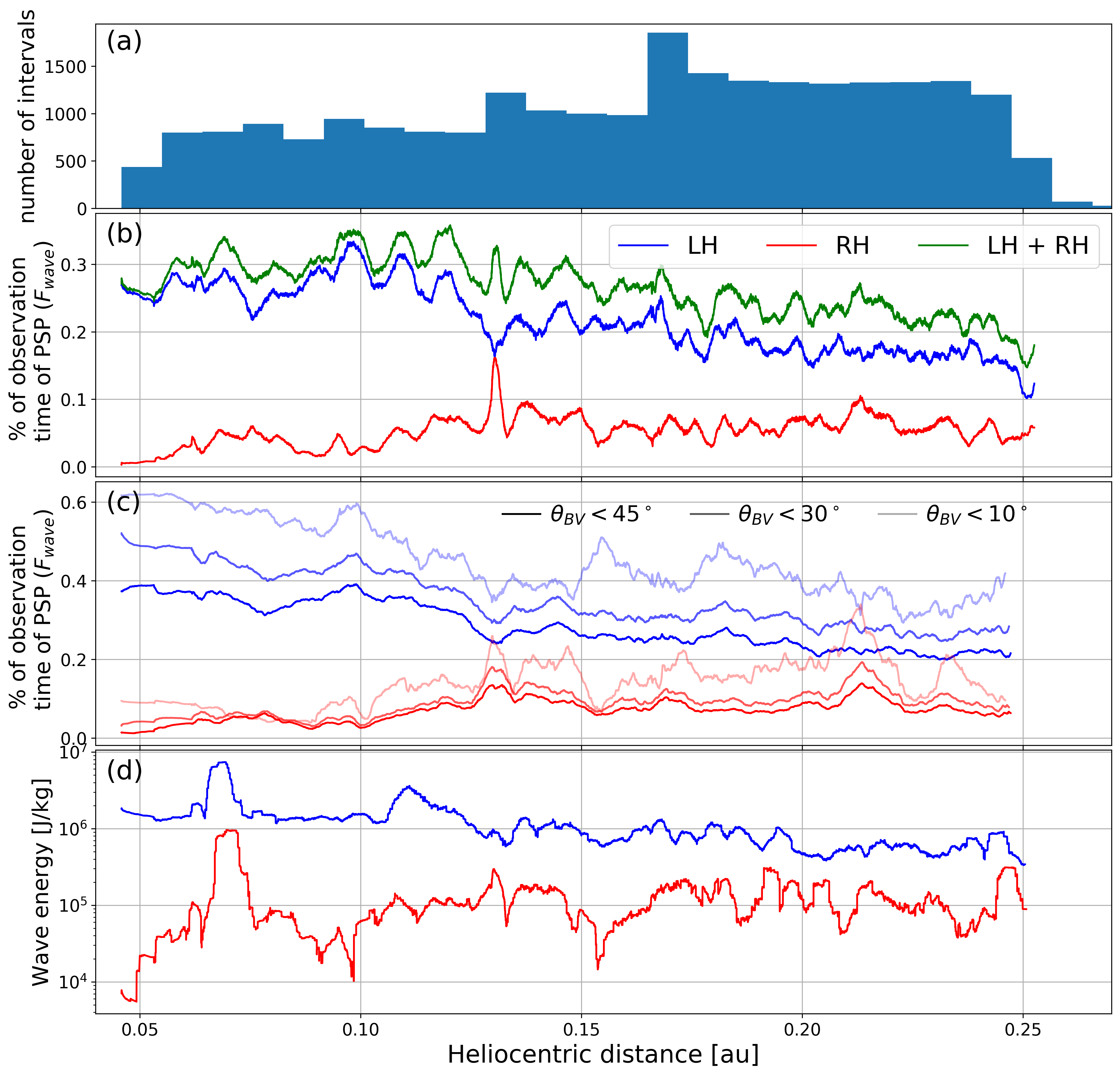}
    \caption{(a) Histogram of PSP observational time with respect to heliocentric distance. 
    Radial profiles of moving means of: 
    (b) fraction of observational time ($F_{wave}$) during which LHW (blue), RHW (red), and both LHW and RHW (green) are observed; (c) $F_{wave}$ of LHW (blue) and RHW (red) when observations whose sampling angle, $\theta_{BV}$ lies below a threshold value; the higher the $\theta_{BV}$ threshold, the more opaque the curve;
    (d) energy in LHW (blue) and RHW (red). 
    LHWs dominate the PSP observations.}
    \label{fig:F_wave_Vs_R_and_wave_energy_Vs_R}
\end{figure*}

Moreover, the fractions of observational time ($F_{wave}$) during which LHWs and RHWs are observed are evaluated for continuous 15-minute intervals for Encounters 1-24. 
In Figure \ref{fig:F_wave_Vs_R_and_wave_energy_Vs_R}, panel (b) shows the moving means of $F_{wave}$ for LHWs (blue), RHWs (red), and both LHWs and RHWs (green) as a function of the heliocentric distance. 
We find that $F^{LH}_{wave}$ increases gradually with decreasing heliocentric distance, reaching up to 30\% within 0.15 au, while $F^{RH}_{wave}$ remains low and slightly decreases with decreasing heliocentric distance (consistent with \cite{Liu_2023} who analyze PSP observations from Encounters 1 - 11). 

Importantly, the evaluated $F_{wave}$ radial profiles have a strong sampling angle ($\theta_{BV}$) bias (see Section \ref{sec:F_wave_Vs_sampling_angle} and Appendix \ref{sec:_meth_wave_theta_anti_corr} for details). 
To account for this bias, we evaluate $F_{wave}$ considering observations whose $\theta_{BV}$ lie below a threshold value. 
Panel (c) illustrates the estimated $F_{wave}$ when $\theta_{BV}$ thresholds vary between $45^\circ-10^\circ$ (decreasing opacity). 
Accounting for sampling angle bias greatly enhances the LHW occurrence. 
The lower the $\theta_{BV}$ threshold, the larger the estimated $F^{LH}_{wave}$ radial profile, which reaches up to 60\% near the Sun, underlining the continuous generation and dissipation of ICWs at these distances. 
On the other hand, $F^{RH}_{wave}$ remains relatively unchanged for all $\theta_{BV}$ threshold values at smaller radial distances. $F^{RH}_{wave}$ increases notably at larger radial distances when $\theta_{BV} < 10^\circ$ possibly because of increased detection of parallel FMWs. Note that observations from Encounters 1 and 2 are not considered in panel (c) because of the unavailability of SPAN observations.
Lastly, the energies of ion-scale waves for all 15-minute intervals are evaluated. 
Panel (d) shows the moving means of the LHW (blue) and RHW (red) energy. 
LHWs predominant at all heliocentric distances.

\section{Dependence of Ion-scale wave observation on the sampling angle} \label{sec:F_wave_Vs_sampling_angle} %, $\theta_{BV}$

\begin{figure*}[hbt!]
    \centering
    \includegraphics[width=0.95\linewidth]{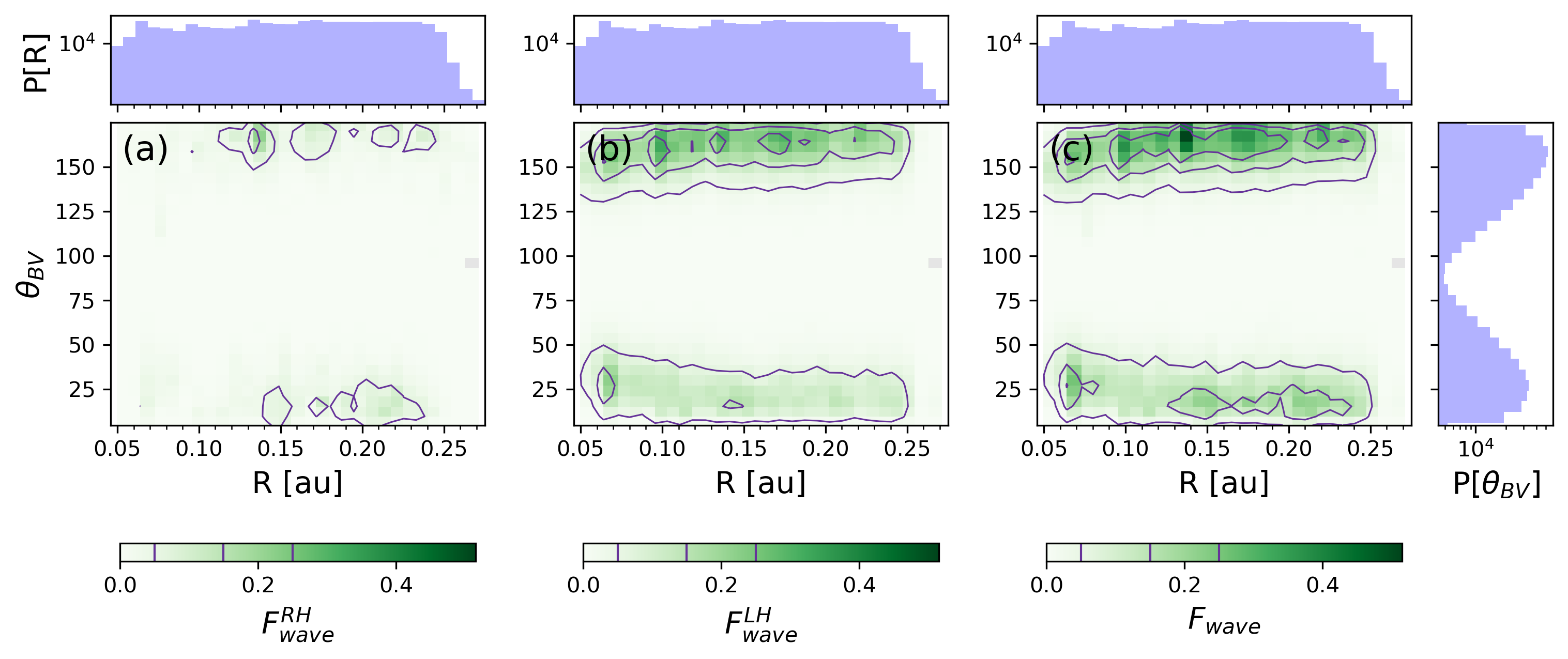}
    \caption{2D histograms of time fraction of wave observation ($F_{wave}$) of (a) RHWs, (b) LHWs, and (c) both LHWs and RHWs as a function of sampling angle ($\theta_{BV}$) and heliocentric distance (R) for Encounters 3-24. 
    The colorbar value in a 2D bin is the bin-mean value of  $F_{wave}$ normalized by a factor of $N_i/N_{max}$, where $N_i$ is the number of data points in the ith bin and $N_{max}$ is the maximum of the set of numbers of data points in all bins. 
    The identical horizontal histograms on the top are of R, and the vertical histogram on the right is of $\theta_{BV}$. The purple isocontours correspond to $F_{wave}$ values of 0.05, 0.15, and 0.25, respectively, as indicated on the colorbars.}
    \label{fig:theta_BV_R_F_wave_2D_hist}
\end{figure*}

Figure \ref{fig:theta_BV_R_F_wave_2D_hist} shows the statistical dependence of $F_{wave}$ on the sampling angle, $\theta_{BV}$, and R. 
We infer that the LHWs are preferentially observed when the spacecraft-frame solar wind velocity is parallel or antiparallel to the local mean \text{B} direction. However, this dependence is less pronounced for RHWs. 

The magnetic field fluctuations observed by PSP are a combination of distinct fluctuations, including ion-scale waves and turbulent fluctuations. 
Parallel propagating ion-scale waves have smaller parallel (with respect to \textbf{B}) wavevectors, whereas turbulent fluctuations have smaller perpendicular wavevectors \citep{Horbury2012_k_aniso}. 
Hence, the characteristics of fluctuations observed by a single spacecraft depend on the sampling angle and the relative energies in the projected wavevectors of waves and turbulence along the sampling direction, i.e., the solar wind velocity vector in the PSP frame. 
\cite{Bowen_2020_ICW} first reported the preferential observation of ion-scale waves by PSP when the local mean \textbf{B} is radial, i.e., $\theta_{BV}$ is close to $0^\circ$ or $180^\circ$ for Encounter 1 observations. 
They hypothesized that this observational bias is due to the anisotropic turbulent fluctuations that are dominant over waves when $\theta_{BV}$ is close to $90^\circ$ overwhelming the signal of the coherent parallel propagating waves. 
Our work extends this inference to all PSP observations. 
The striking dependence of $F_{wave}^{LH}$ on $\theta_{BV}$ suggests that LHWs probably occur for longer durations, although they appear/disappear in observations depending on the value of $\theta_{BV}$. 
The weak correlation of $F_{wave}^{RH}$ with $\theta_{BV}$ suggests that RHWs are a mixture of parallel and oblique propagating waves.
This dependence of $F_{wave}^{LH/RH}$ on $\theta_{BV}$ is consistent with the evaluated Pearson correlation coefficients between the two parameters for LHWs and RHWs as illustrated in Figure \ref{fig:wave_theta_corr_example} and panel (f) of Figure \ref{fig:E12_encounterwise_trends}.
Observations from Encounters 1 and 2 are not included from here on as reliable SPANi observations are unavailable during these encounters.

\section{Waiting and Residence times of Ion-scale waves} \label{sec:waiting_residence_times}

\begin{figure}[hbt!]
    \centering
    \includegraphics[width=0.95\linewidth]{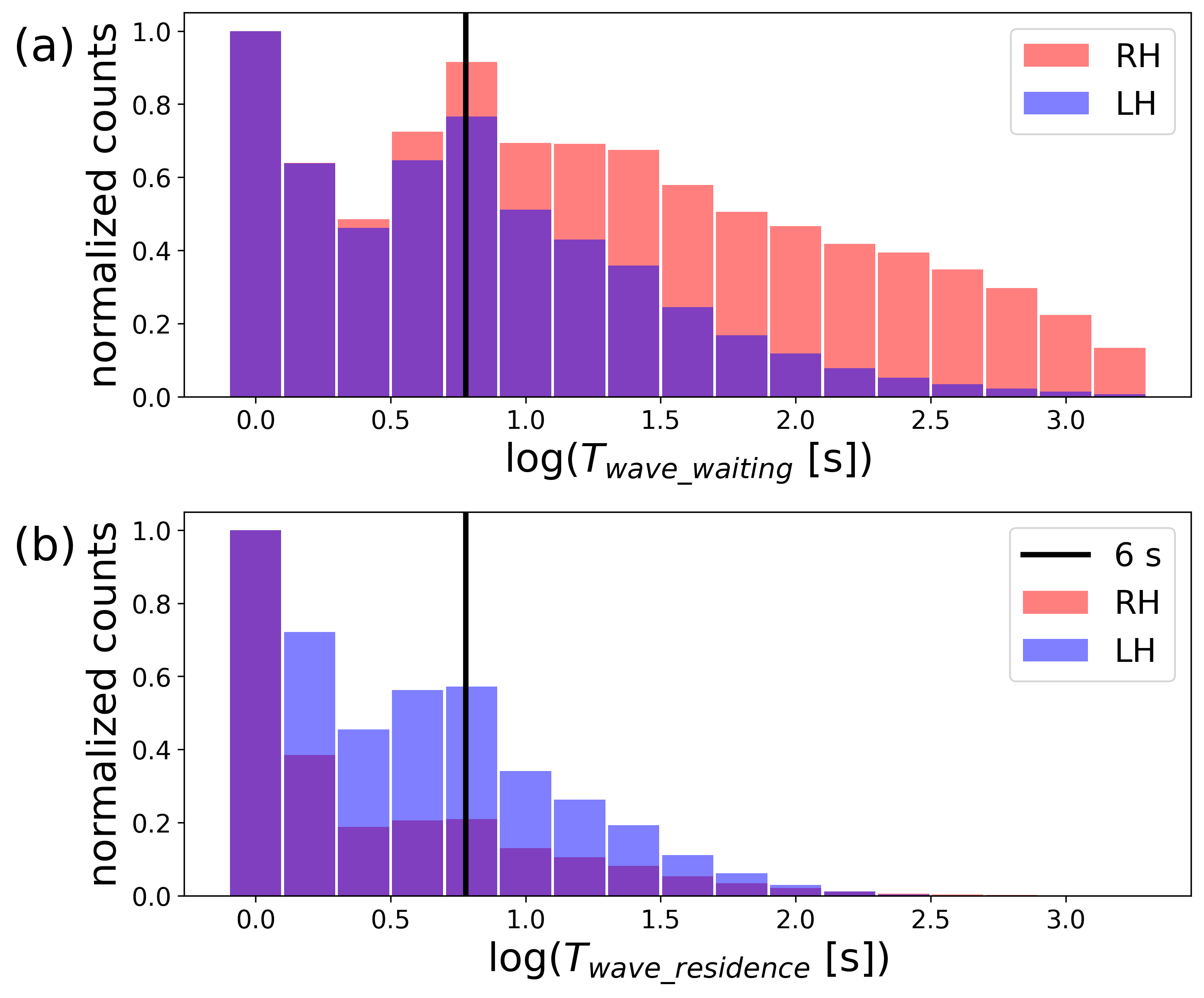}
    \caption{Histograms of (a) waiting times and (b) residence times of ion-scale waves for Encounters 3-24. The histograms are normalized such that the maximum value is 1. The black line corresponds to the timescale for $\theta_{BV}$ to fluctuate by $\sim 15^\circ$, see Fig.~\ref{fig:theta_BV_timescale_dist}, sufficient to obscure or reveal parallel propagating waves.}
    \label{fig:wave_waiting_residence_times}
\end{figure}

Figure \ref{fig:wave_waiting_residence_times} illustrates the distribution of waiting and residence times for ion scale waves, following \cite{Dudok_de_Wit_2020}. 
Here, \emph{waiting time} refers to the time duration between two consecutive wave events, and \emph{residence time} is the time width of wave events, where a wave event is a continuous time patch during which a wave is identified without a break. 
Note that in the waiting and residence time distributions, there is a peak around 1 NYs, partly because it is the maximum time resolution of the wave repository, and all waiting/residence times $<$ 1 NYs have not been resolved and have instead been downsampled to 1 NYs. 
Moreover, waves with lower amplitudes are often interrupted by turbulent fluctuations. 
Furthermore, the time-averaging of $\sigma(s,t)$ employed to identify these waves (see Appendix A of \cite{Niranjana_2024}) causes the value of $\sigma$ to drop below the identification threshold in many time bins within a weak wave event. 
Thus, weaker waves are identified as several transient wave events, even though they might be single long-lived waves, leading to the peak near the lowest waiting and residence times. 
The secondary peak at $\sim$6 s (black line) is likely caused by the rotation of the sampling angle. 
As shown in Figure \ref{fig:theta_BV_timescale_dist} and described in Appendix \ref{sec:_meth_wave_theta_anti_corr}, 6 seconds is approximately the timescale over which $\theta_{BV}$ fluctuates by $\sim 15^\circ$. 
Fluctuations in the sampling angle of this amplitude are sufficient to obscure or reveal the observation of parallel propagating LHWs and RHWs. 
%We argue that it is reasonable to expect a parallel wave to appear or disappear in observations when the sampling angle rotates by $15^\circ$.

The RHW waiting times are significantly longer than the LHW waiting times, while the RHW residence times are much shorter than the LHW residence times, consistent with our inference that RHWs occur less frequently and in bursts. 
The longer waiting times of RHWs potentially correspond to the duration between consecutive measurements near the HCS or other large scale structures capable of driving a proton beam. 

The LHW waiting and residence time distributions are similar. 
This similarity is consistent with the scenario in which LHWs occur continuously over long durations, and they appear/disappear in observations as the sampling angle ($\theta_{BV}$) fluctuates. 
The $\theta_{BV}$ fluctuates over a broad band of time scales due to turbulence. 
The core distributions of the LHW waiting and residence times around the secondary peak are characteristic of the set of these $\theta_{BV}$ fluctuation time scales. 
The waiting and residence times statistics presented here are consistent with panel (a) of Figure \ref{fig:E12_encounterwise_trends}.

\section{Temperature Anisotropy Instabilities And Observed Ion-scale Waves} \label{sec:F_wave_Vs_T_p_aniso}

In Figure \ref{fig:beta_p_T_p_aniso_F_wave_2D_hist}, the time fraction of wave observation ($F_{wave}$) is plotted as a function of $T_{\perp,p}/T_{\parallel,p}- \beta_{\parallel,p}$ space and compared with the temperature anisotropy thresholds (Equation \ref{eq:T_p_aniso_thresh}). 
We find that LHWs are most frequently observed near the ion cyclotron (A/IC) threshold with a smaller increase near the parallel firehose (pFH) instability threshold.
RHWs are more frequently observed near the parallel firehose instability threshold. This dependence of wave observation on $T_{\perp,p}/T_{\parallel,p}- \beta_{\parallel,p}$ space is consistent with \cite{Liu_2025_waves_Tp_aniso_E4_E11}, who focused on data from Encounters 4 through 11.
We infer that temperature anisotropy instabilities play an important role in constraining the non-Maxwellian features of proton VDFs. 

Figure \ref{fig:long_beta_p_T_p_aniso_F_wave_R_binned} shows 2D histograms of $F_{wave}$ in the $T_{\perp,p}/T_{\parallel,p}- \beta_{\parallel,p}$ space as a function of heliocentric distance, $R$. 
We find that very close to the Sun (R $<$ 0.09 au), LHWs are very frequently observed near the A/IC threshold, and RHWs are not observed at these distances. 
As we move away from the Sun, between 0.09 and 0.18 au, the occurrence of LHWs near the A/IC threshold gradually decreases, and instead they are increasingly seen near the pFH threshold. 
Moreover, RHWs are also observed near the pFH threshold at these intermediate radial distances. 
At distances further from the Sun (R $>$ 0.18 au), LHWs rarely occur near the A/IC threshold and are less frequently seen near the pFH threshold. 
Moreover, RHWs are observed near the pFH threshold, and their occurrence gradually decreases with increasing R. The radial evolution of PSP observations in the $T_{\perp,p}/T_{\parallel,p}- \beta_{\parallel,p}$ space follow a similar trend to that of $F_{wave}$ (right panels in Figure \ref{fig:long_beta_p_T_p_aniso_F_wave_R_binned} )  and is consistent with previous works that have analyzed observations from various spacecrafts \citep{Maruca_2011_T_p_aniso, Matteini_2012, Huang_2025}.
Additionally, the region occupied by observations in the $T_{\perp,p}/T_{\parallel,p}- \beta_{\parallel,p}$ space (green-shaded region) becomes narrower with increasing R, which is consistent with the instabilities triggering for smaller deviations of $T_{\perp,p}/T_{\parallel,p}$ from unity when $\beta_{\parallel,p}$ values are higher.

However, waves are observed away from the thresholds for a significant duration of time. 
The A/IC instability is triggered by a high proton core $T_{\perp,p}/T_{\parallel,p}$, and the pFH instability is triggered by excess parallel pressure irrespective of the contributions from the proton VDF components. 
Thus, the occurrence of a significant number of ion-scale waves away from the A/IC and pFH thresholds implies that these waves cannot be driven by the temperature anisotropy of a single proton component alone, but by other sources of free energy, such as proton beams and alpha particles. Moreover, the instability thresholds assume that the wave amplitude remains small, which is only true during the initial stages of the instability growth.
See Section \ref{sec:conclusions} for a discussion about quasilinear (QL) relaxation and models and theories that describe ICW heating using QL diffusion.

\begin{figure*}[hbt!]
    \centering
    \includegraphics[width=0.95\linewidth]{figures/beta_p_Tp_aniso_all_encounters.png}
    \caption{2D histograms (similar to Figure \ref{fig:theta_BV_R_F_wave_2D_hist}) of time fraction of wave observation ($F_{wave}$) of (a) RHWs, (b) LHWs, and (c) both LHWs and RHWs as a function of proton temperature anisotropy ($T_{\perp,p}/T_{\parallel,p}$) and parallel proton plasma beta ($\beta_{\parallel,p}$) for Encounters 3-24. 
    The thresholds of ion cyclotron (red), parallel firehose (dark blue), mirror (light blue), and oblique firehose (orange) are plotted. 
    The identical horizontal histograms on the top are of $\beta_{\parallel,p}$ and the vertical histogram on the right is of $T_{\perp,p}/T_{\parallel,p}$. 
    The purple isocontours correspond to $F_{wave}$ values of 0.05, 0.15, and 0.25, respectively, as shown on the colorbars.}
    \label{fig:beta_p_T_p_aniso_F_wave_2D_hist}
\end{figure*}

\begin{figure*}[hbt!]
    \centering
    \includegraphics[width=0.70\linewidth]{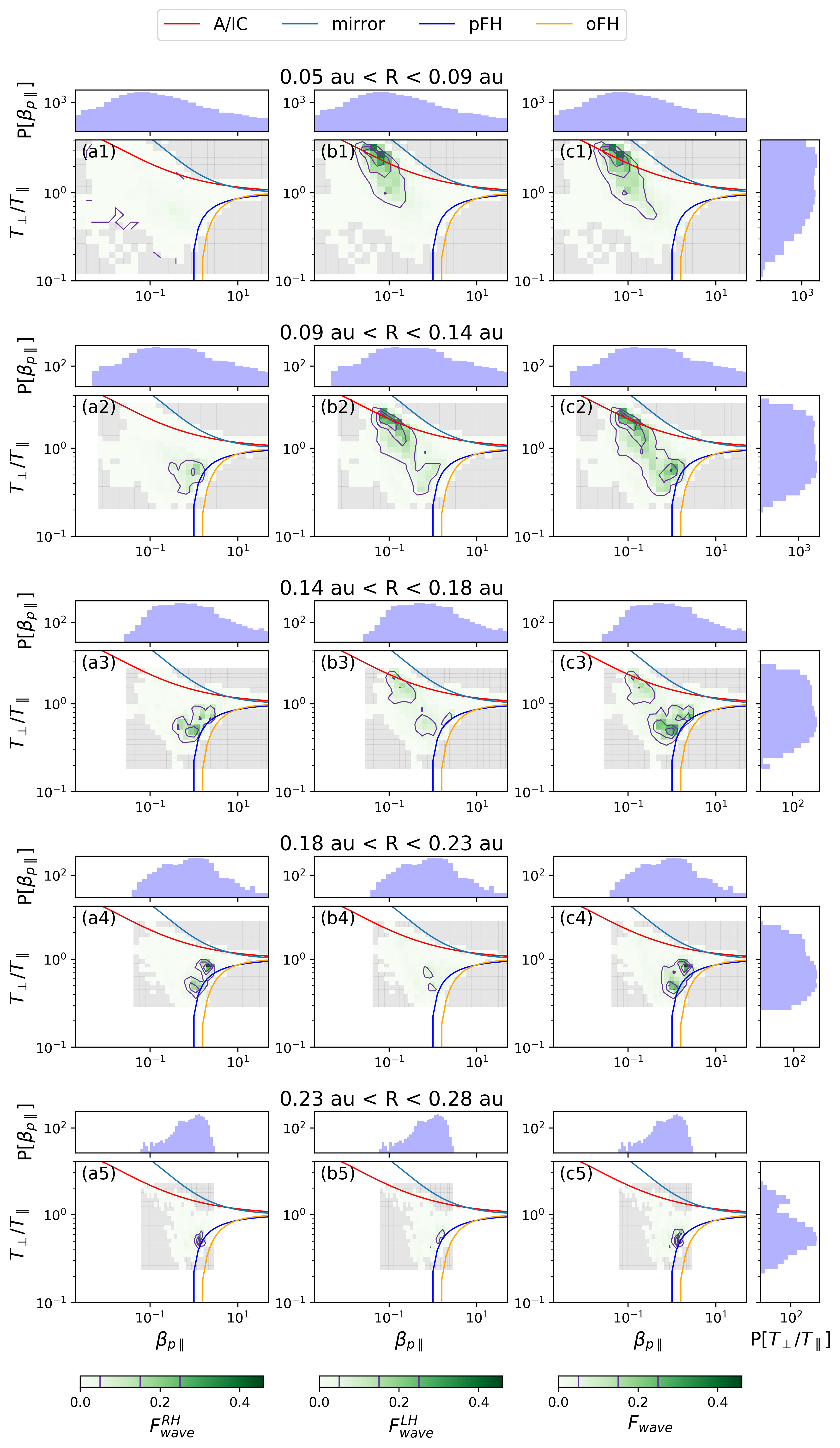}
    \caption{2D histograms of $F_{wave}$ for LHWs, RHWs, and both LHWs and RHWs in $T_{\perp,p}/T_{\parallel,p}- \beta_{\parallel,p}$ space which are similar to Figure \ref{fig:beta_p_T_p_aniso_F_wave_2D_hist}, but binned with respect to heliocentric distance, R for Encounters 3-24. Note that for each radial bin, the colorbar is normalized using $N_{max}$ similar to Figure \ref{fig:theta_BV_R_F_wave_2D_hist}. However, the number of observations and $N_{max}$ vary over different radial bins depending on the observation time of PSP and the goodness of SPANi FOV at various radial distances.}
    \label{fig:long_beta_p_T_p_aniso_F_wave_R_binned}
\end{figure*}

\section{Proton beam instabilities and observed RHWs} \label{sec:F_wave_Vs_q_par}

\begin{figure*}
    \centering
    \begin{minipage}{0.425\textwidth}
        \includegraphics[width=0.9\linewidth]{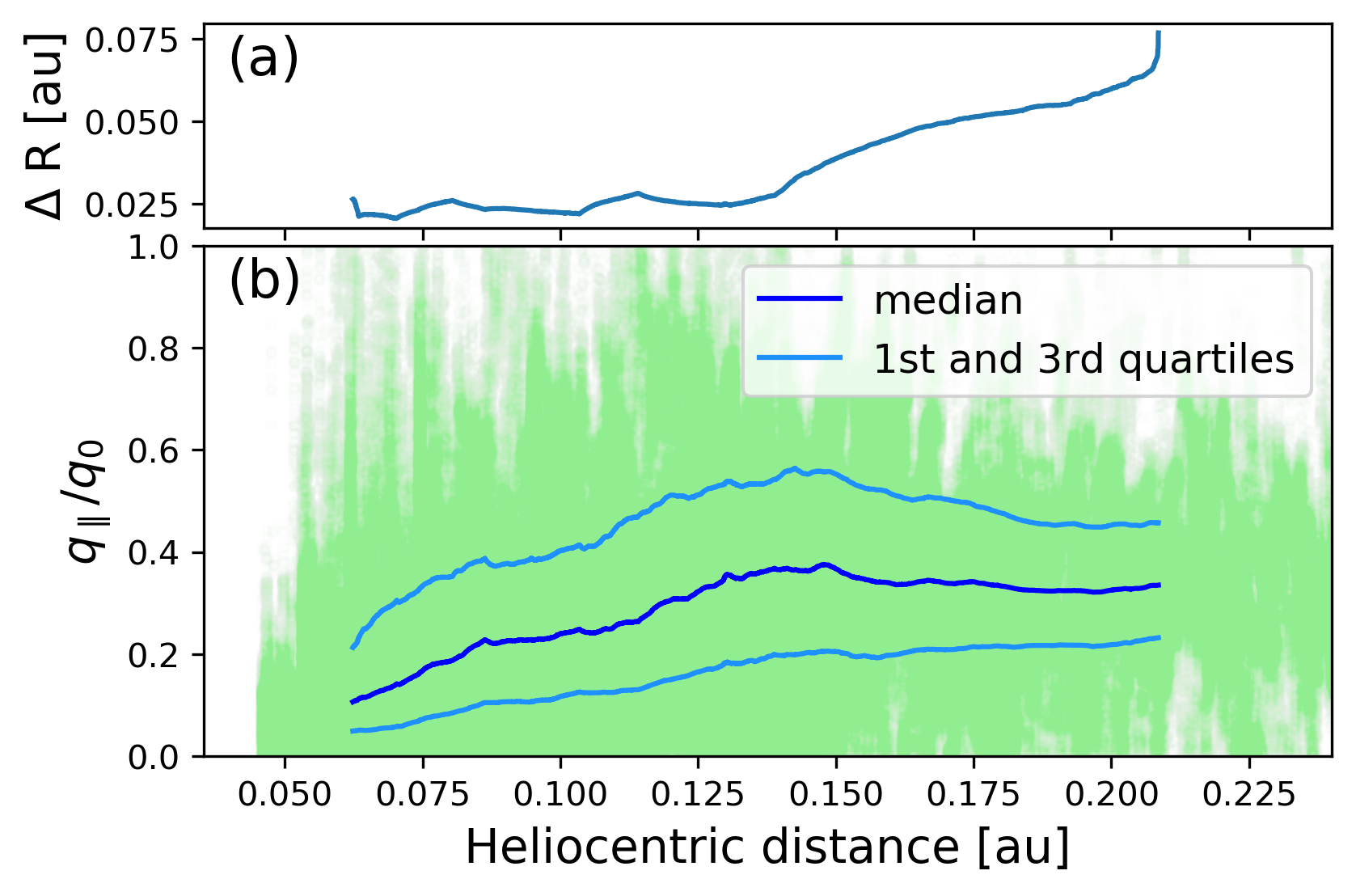}
        \caption{(b) The normalized proton parallel heat flux ($q_{\parallel,p}/q_{0,p}$, green) for Encounters 3 - 24 as a function of heliocentric distance, along with the first three moving quartiles (light and dark blue curves). 
        The quartiles are evaluated over moving windows with a width $\Delta R$ necessary to encompass $10^6$ observations, with the corresponding widths in radial distance shown in (a).}
        \label{fig:q_par_Vs_R}
    \end{minipage} \hfill \begin{minipage}{0.525\textwidth}
        \centering
        \includegraphics[width=1\linewidth]{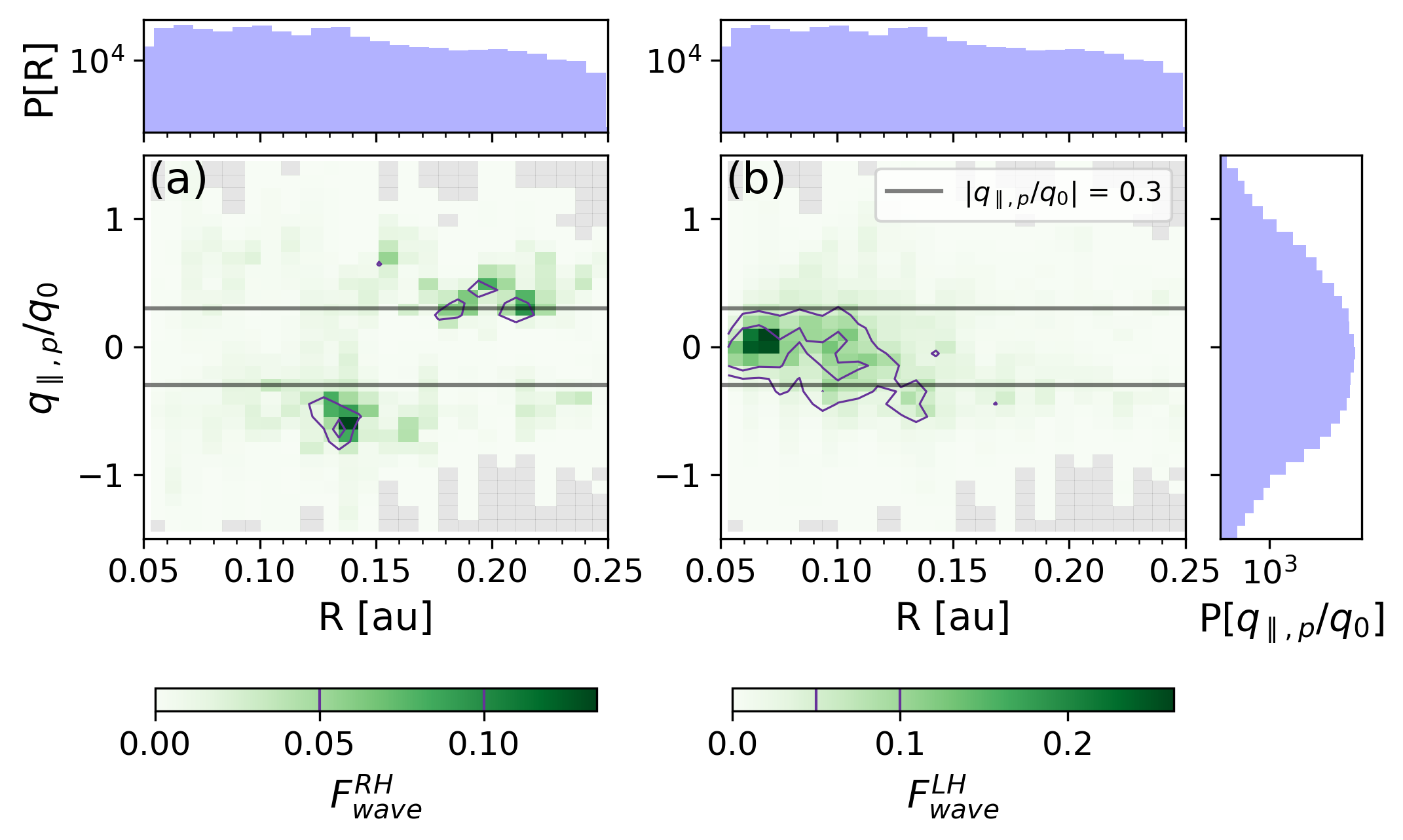}
        \caption{2D histograms (similar to Figure \ref{fig:beta_p_T_p_aniso_F_wave_2D_hist}) of time fraction of wave observation ($F_{wave}$) of (a) RHWs and (b) LHWs as a function of normalized proton parallel heat flux ($q_{\parallel,p}/q_{0,p}$) and heliocentric distance (R) for encounters 3 - 24. 
        The purple isocontours correspond to $F_{wave}$ of 0.05 and 0.1. 
        The black lines correspond to the inferred threshold of $|q_{\parallel,p}/q_{0,p}|$ = 0.3, beyond which RHWs are likely to be driven unstable by the proton beam.}
        \label{fig:q_par_R_F_wave_2D_hist}
    \end{minipage}
\end{figure*}

In this work, we quantify the presence of proton beams using the proton parallel heat flux as an effective proxy. 
We evaluate the normalized proton heat flux, $q_{\parallel,p}/q_{0,p}$, for SPANi observations (as described in Section \ref{sec:meth_PSP_SPAN}) from Encounters 3 - 24, shown in Figure \ref{fig:q_par_Vs_R}  as a function of the heliocentric distance (green in panel (b)) along with the first three moving quartiles (light and dark blue curves in panel (b)). 
The variation of the median heat flux with respect to the heliocentric distance changes at $\sim$0.14 au, a distance coincident with the mean Alfv\'en surface \citep{Chhiber_2022_Alfven_surface, Badman_2025_Alfven_surface}. 
The median heat flux decreases below $R \sim 0.14$ au and is constant for larger distances.

\begin{figure*}
    \centering
    \includegraphics[width=0.95\linewidth]{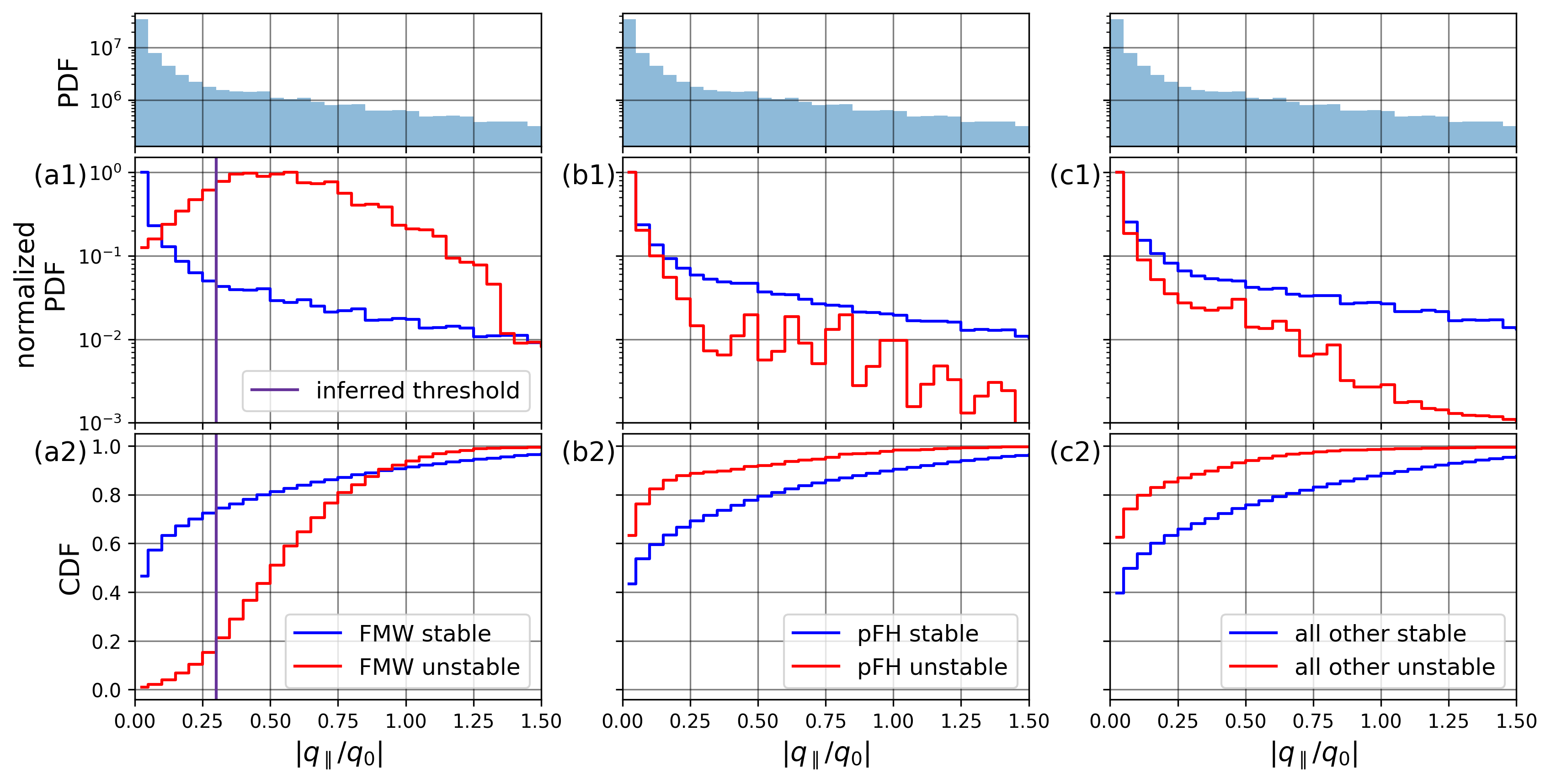}
    \caption{The normalized histograms of $|q_{\parallel,p}/q_{0,p}|$ values of stable and unstable intervals predicted by \texttt{SAVIC} for (a1) FMW instability, (b1) parallel firehose instability, and (c1) all other instabilities, as well as their corresponding cumulative distributions (lower panels). 
    The identical histograms in the top panel are of $|q_{\parallel,p}/q_{0,p}|$ values corresponding to $\mathcal{P}$. 
    The purple line is the inferred threshold on $|q_{\parallel,p}/q_{0,p}|$ above which FMWs are likely driven unstable by proton beams. 
    The histograms in the center panel are normalized such that their maximum value is 1.}
    \label{fig:SAVIC_FMW_q_par_thresh}
\end{figure*}

Figure \ref{fig:q_par_R_F_wave_2D_hist} shows the 2D histograms of the time fraction of wave observation ($F_{wave}$) as a function of $q_{\parallel,p}/q_{0,p}$ and R. 
We infer that, generally, RHWs are associated with higher heat flux, while LHWs are associated with lower heat flux. 
This observed correlation is consistent with the RHWs being parallel and oblique FMWs, and the LHWs being ICWs. 
We identify an observational threshold of $|q_{\parallel,p}/q_{0,p}| \sim 0.3$ beyond which RHWs are driven unstable. 

To further validate the threshold estimate, we use the \texttt{SAVIC} machine learning algorithm. 
A multi-parameter space is generated by varying the set of parameters $\mathcal{P} = \left[\beta_{\parallel,pc}, \dfrac{T_{\perp,pc}}{T_{\parallel,pc}}, \dfrac{T_{\parallel,pc}}{T_{\parallel,pb}}, \dfrac{T_{\perp,pb}}{T_{\parallel,pb}}, \dfrac{n_{pb}}{n_{pc}}, \dfrac{v_{bc}}{v_A} \right]$ uniformly distributed over the set of ranges ($10^{-2}$-10,  $10^{-1}$-10, 0.5 - 2, 0.1 - 2, $10^{-2}$ - 0.5, 0-2), respectively, based on the observed values in the inner heliosphere. 
The value of $q_{\parallel,p}/q_{0,p}$ and the most unstable mode predicted by \texttt{SAVIC} are calculated for each set of plasma parameters in $\mathcal{P}$ (see Appendix \ref{sec:q_par_bi_Maxwellian} for the expression of $q_{\parallel,p}/q_{0,p}(\mathcal{P})$). 
Figure \ref{fig:SAVIC_FMW_q_par_thresh} shows the normalized histograms of the $|q_{\parallel,p}/q_{0,p}|$ values of the stable and unstable intervals for FMW (panel (a1)), parallel firehose (panel (b1)), and all other instabilities excluding FMW and parallel firehose instabilities (panel (c1)) as predicted by \texttt{SAVIC} along with their corresponding cumulative distributions (lower panels). 
We infer that only for FMW instability, the fraction of unstable (stable) intervals increases (decreases) with increasing $|q_{\parallel,p}/q_{0,p}|$. 
Thus, the observed RHWs are consistent with parallel and oblique propagating FMWs that are driven unstable by proton beams, represented by stronger heat fluxes.

The observationally inferred threshold of 0.3 on $|q_{\parallel,p}/q_{0,p}|$ (purple line in panels a1 and a2) is consistent with \texttt{SAVIC} predictions. 
A majority of FMW unstable intervals (85\%) occur above the threshold, and most of the FMW stable intervals ($\sim$72\%) occur below the threshold. 
A possible critique of the proposed threshold is that the above-mentioned percentages do not capture all of the unstable FMWs. 
However, the stability of FMWs in the presence of a proton beam has a complex dependence on proton VDF characteristics, such as the number of core and beam resonant particles, the core anisotropy, and the gradient of VDF magnitude along specific contours, and reducing this complex dependence to a single parameter introduces error. 
Moreover, the parameter space $\mathcal{P}$ used to predict the instabilities is uniformly distributed and is not representative of the observed distributions. 
Furthermore, \texttt{SAVIC} predictions have an error of $\sim 2.5\%$. 
Taking into account these caveats, we argue that the proposed threshold on $|q_{\parallel,p}/q_{0,p}|$ for the FMW instability is a qualitatively accurate boundary in this high-dimensional space.
%\begin{figure} 
    
%\end{figure}

\section{Conclusions} \label{sec:conclusions}

A striking feature of PSP observations is the abundant observation of circularly polarized ion-scale waves \citep{Bowen_2020_ICW, Liu_2023}. 
The observed trends of ion-scale waves are signatures of the interplay between various turbulent dissipation mechanisms and instabilities that drive these waves. 
We undertook a mission-wide statistical survey of observed ion-scale waves and created a repository of these waves (\cite{Niranjana_2025_Zenodo}). The survey confirms and statistically extends the findings of several previous smaller case studies.
We find that LHWs are frequently observed, and the time fraction of their observation gradually increases closer to the Sun, reaching up to $\sim$30\% (consistent with \cite{Bowen_2020_ICW} and \cite{Liu_2023}).
The observation of these LHWs strongly depends on the sampling angle (consistent with \cite{Bowen_2020_ICW}), and their waiting times are shorter, suggesting that they probably occur for longer durations than they are observed. Upon accounting for the sampling angle bias, we find that LHWs can occur during 60\% of PSP observational time at closer distances to the Sun for intervals with $\theta_{vB}<10^\circ$.
We find that the observed LHWs are generally consistent with anti-sunward parallel-propagating ICW wave storms (consistent with \cite{Bowen_2020_outward_ICWs}) that occur continuously over extended periods of time (up to several days). 
We further find that the slopes of turbulent energy spectra at kinetic scales are steeper during LHW storms. 
Thus, the observed LHW storms are signatures of processes such as stochastic heating \citep{Chandran_2010_stochastic_heating} and helicity barrier \citep{Squire_2022_helicity_barrier} that are associated with enhancements in local temperature anisotropy which is constrained by the temperature anisotropy instabilities through the generation of ICWs. The driven ICWs are associated with steepening of turbulent spectra at dissipation scales \citep{Bowen_2024_extended_cylotron_heating}. 

This work provides observational constraints on theories that describe the interplay between processes that are mediated by ICWs. 
Dissipation of ICWs has been considered a likely mechanism to heat the solar wind (see review in \cite{Hollweg_Isenberg_2002_wave_heating}). 
While previous studies \citep{Hollweg_1973_wave_heating_model, Isenberg_1982_wave_heating_model, Marsch_1982_ICW_heating_model} considered fluid protons, \cite{Isenberg_2001_kinetic_shell_1, Isenberg_2001_kinetic_shell_2, Isenberg_2004_kinetic_shell_3} developed a \emph{kinetic-shell model} to describe the ICW-proton resonance interaction and the resulting heating by quasilinear (QL) diffusion \citep{Kennel_Engelmann_1966, Kennel_petschek_1966}. 
They argued that due to interaction between parallel ICWs and the proton VDF, $f_p$, the latter would quickly relax toward a state of marginal stability, preventing further damping of ICWs, resulting in weak proton heating. 
They concluded that ICW heating is insufficient to drive the fast solar wind. \cite{Chandran_2010_oblique_par_ICWs} provided an alternate explanation arguing that if ICWs with a range of propagation angles (with respect to \textbf{B}) were continuously generated, then parallel ICWs would be amplified, resulting in $f_p$ to be marginally stable to parallel ICWs, but the oblique ICWs would continue to damp, resulting in a more effective proton heating. 

\cite{Meyrand_2021} and \cite{Squire_2022_helicity_barrier} proposed a novel \emph{helicity barrier} mechanism in which an imbalanced turbulent cascade would channel the energy to oblique ICWs due to a barrier that arises at ion scales because of the conservation of magnetic helicity, which prevents the cascade to KAWs. They further demonstrated that the oblique ICWs would quickly damp onto protons, leading to the growth of parallel ICWs, which cool the plasma. Moreover, the oblique and parallel ICWs generated by the helicity barrier are outward propagating as determined by the direction of imbalance, which is consistent with observations. 

Using PSP observations, \cite{Mcintyre_2025_helicity_barrier} found evidence that the helicity barrier plays a significant role in near-Sun solar wind. However, contrary to the predictions of \cite{Squire_2022_helicity_barrier},  \cite{Bowen_2024_extended_cylotron_heating} estimated that parallel ICWs heat the plasma instead of cooling it. Building upon the previous studies, \cite{Yerger_2024_AGU} have developed a promising comprehensive model for ICW heating by considering WKB evolution of large-scale Alfv\'en waves, their cascade to ion scales, subsequent channeling of energy to oblique ICWs due to helicity barrier, and the consequent driving of anti-sunward propagating parallel ICWs by \emph{quasilinear focusing}. Their model predicts a continuous emission and absorption of parallel ICWs termed as \emph{cyclotron breaking}, which is consistent with our findings.

%The LHW storms thus mediate the local transport of free energy associated with temperature anisotropy and the resulting solar wind heating.

On the other hand, RHWs are less frequently observed \citep{Verniero_2020_waves_proton_beams, Liu_2023}. 
The time fraction of their observation remains low and decreases closer to the Sun. 
Their observation weakly depends on the sampling angle, and they have longer waiting times between observations. 
The observation of RHWs is well correlated with an increase in proton parallel heat flux. 
The observed RHWs are generally consistent with bursty parallel and oblique propagating FMWs, and they mediate the free energy associated with proton beams. 
Moreover, we identify a threshold on normalized heat flux above which the FMWs are likely to be driven unstable by proton beams. 
Furthermore, cross helicity decreases when RHWs are observed, possibly because of proximity to the heliospheric current sheet, where the observed proton beams are generated \citep{Phan_2022_p_beams_reconnection}.

%A useful extension of this work is to investigate the dependence of wave observation on the distance from the Alfv\'en surface. 
%It will be further useful to estimate the possible dissipation rates of observed ICW storms via cyclotron resonance. 
%In the future, we hope to study the possible dependence of the observed RHWs and proton beams on the proximity to HCS as well as the possibility of beam generation by Landau damping of compressive waves \citep{Li_2010_p_beam_by_KAW, Voitenko_2015_p_beam_KAW_turb}. 

In this work, we have not taken into account the impact of alpha particles and we used the protons as a single population rather than distinguishing them into core and beam populations.
Neglecting the multiple ion components can significantly impact the derived dispersion relation \cite{Klein:2021,Martinović_2025}.
Importantly, single component models predict waves that propagate both along and opposite to local \textbf{B}, which is in contradiction with observations where LHWs are generally anti-sunward propagating. 
A more accurate analysis of the observed waves by distinguishing VDFs into proton cores and beams and including alpha particles will be a topic for future work. 
The observed VDFs also deviate from bi-Maxwellian distributions commonly used in linear theory.
Such deviations can cause significant differences in the predictions of growth and damping rates \citep{Walters_2023,Klein:2025-ALPS}.

Furthermore, the instability thresholds used in this work are evaluated using linear plasma dispersion solutions, which assume small-amplitude waves that do not change the bi-Maxwellian VDF. 
However, wave generation due to an instability is a non-linear process where the waves grow and back-react on the VDF.
While the instability thresholds provide a simple model to compare with observations, they do not account for the VDF changes due to a growing instability. 
Quasilinear (QL) diffusion theory accounts for a slow change in f due to a growing instability and provides a better description of the latter. 
The QL theory predicts that f changes such that its gradients along specific diffusion contours vanish. 
In observations, the magnitude f has indeed been found to be constant along the QL diffusion contours \citep{Marsch_Tu_2001_QL_contours, He_2015_QL_contours, Bowen_2022_QL_contours}.

\begin{acknowledgments}

The Parker Solar Probe was designed and built and is now operated by the Johns Hopkins Applied Physics Laboratory as part of NASA's Living with a Star (LWS) program (contract NNN06AA01C). 
Support from the LWS management and technical team has played a critical role in the success of the Parker Solar Probe mission.
N.S. and K.G.K. were additionally supported by NASA grant 80NSSC24K0171.

We acknowledge LESIA (Laboratoire d Etudes Spatiales et Instrumentation en Astrophysique), Observatoire de PARIS, CNRS (Centre National de la Recherche Scientifique) for SQTN data. 
We thank Orlando Romeo for assistance in applying the FOV criteria to SPANi observations. 
We thank Michael Terres and Srijan Das for their help in evaluating SPANi proton VDFs.
\end{acknowledgments}

%\clearpage

\appendix

\section{Methodology} \label{sec:methodology}

In this section, the methodology used in this work is described in more detail.

\subsection{PSP/FIELDS Data} \label{sec:meth_PSP_FIELDS}
We divide PSP/FIELDS flux gate magnetometer (MAG) magnetic field (\textbf{B}) observations \citep{Bale2016} from Encounters 1-24 into continuous 15-minute non-overlapping intervals following the procedure described in Appendix A.1. of \cite{Niranjana_2023} (SKM23 from here on). 
For each interval, we identify and remove the reaction wheel noise (see Appendix B.1 of SKM23), downsample the B time series to frequencies at which the observed \textbf{B} energy spectra hit the MAG noise floor \citep{Bowen_noise}, and evaluate the spectrum of spacecraft-frame polarization,
\begin{equation}
    \sigma(s,t) = \frac{-2 \text{Im} \left( W_X(s,t)W_Y^*(s,t) \right)}{|W_X(s,t)|^2 + |W_Y(s,t)|^2}
\end{equation}
in a scale-sensitive local magnetic field aligned coordinate system (\textbf{X,Y,Z}) using Mortlet wavelet transforms (\cite{Bowen_2020_ICW}, see Appendices B.3. and B.4. of SKM23).
Here $W_{XY}(s,t)$ are Mortlet wavelet transforms of the magnetic field components in the plane perpendicular to the local magnetic field, which is parallel to \textbf{Z}, and \textit{s} is the inverse frequency. 
A $\sigma$ value of +1 (-1) represents ion-resonant/left-handed (electron-resonant/right-handed) polarization in the spacecraft frame. 
We then identify the time and frequency bins where persistent high values of circular polarization corresponding to ion-scale waves are observed by applying a time-smoothing routine and selecting frequency-time regions with $|\sigma| > 0.7$ as described in Appendix A of \cite{Niranjana_2024} (SKM24 from here on) and evaluate the energy spectra of ion-scale waves ($|\tilde{B}(f)|^{2^{(LH/RH)}}_{wave}$). 
We further identify time and frequency bins where $|\sigma| <= 0.7$ corresponding to turbulence and evaluate turbulent energy spectra ($|\tilde{B}(f)|^2_{turb}$) over the 15-minute intervals. 
\footnote{Note that in this work, to be consistent with standard practices, we reverse the color code of LH and RH polarizations compared to SKM24 and \cite{Bowen_2020_ICW}.} 

The identified wave spectra are stored with a time cadence of 1 New York second (NYs), where 1 NYs $\sim$0.874 s. 
Upon repetition of the ion-scale wave identification routine for all 15-minute intervals, a mission-wide ion-scale wave repository is evaluated [ZENODO LINK TO BE ADDED]. 
Using the wave spectra repository, \textit{wave\_presence\_index (LH/RH)}, the presence or absence of ion-scale waves in a time bin is evaluated as a function of time with a time bin width of 1 NYs. 
The value of \textit{wave\_presence\_index (LH/RH)} at a time bin is: 0 if a LHW/RHW is not observed at that time bin; 1 if a LHW/RHW is observed in at least one frequency bin at that time bin. 
The time fraction of ion-scale wave observation, $F_{wave}^{(LH/RH)}$, is then evaluated at a required cadence, $\tau$, greater than 1 NYs by binning the \textit{wave\_presence\_index (LH/RH)} with bin width = $\tau$ and evaluating the ratio of duration of the observed waves within the bin to the bin width.

\revise{Furthermore, using the turbulent energy spectra that is evaluated upon removing wave signatures, the steepest slope in the dissipation scales, $\alpha_{max}$, is estimated. Here, the dissipation range is assumed to be the frequency band between 0.1 Hz and the frequency where the observed spectra hit the MAG noise floor.} The turbulent energy spectrum is interpolated at 500 frequency points, and slopes of the energy spectra over a moving window of width 50 points are evaluated. 
$\alpha_{max}$ is assigned the minimum value of the evaluated slopes. \revise{Note that there could be an error in evaluating the steepest slope if the spectral steepening occurs at frequencies higher than the frequency where observed spectra hit the MAG noise floor, but it usually is not the case, especially at closer radial distances.}

\subsection{PSP/SPANi Data} \label{sec:meth_PSP_SPAN}
We use PSP/SPANi \citep{Livi_2022} L2 data to evaluate proton VDFs, their moments, and non-Maxwellian features. 
Initially,  we evaluate the proton VDFs using the differential energy flux, EFLUX [$\text{eV}/\text{cm}^2$-s-ster-eV] parameter, which is a function of ENERGY [eV], THETA [rad], and PHI [rad] bins. 
The discrete values of VDF are evaluated as
\begin{align}
     f_p(\mathbf{v})& [(\text{km/s})^{-3} \text{cm}^{-3} \text{ster}^{-1}] = \frac{EFLUX \times mass_p^2 \times 10^{-5}}{2 \times ENERGY^2} \nonumber \\
     &v_x [\text{km/s}] = |v| \times \cos(PHI) \times \cos(THETA) \nonumber\\
&v_y[\text{km/s}] = |v| \times \sin(PHI) \times \cos(THETA) \nonumber\\
 &v_z[\text{km/s}] = |v| \times \ \ \ \ \ \ \ \ \ \ \ \ \ \ \ \ \ \sin(THETA) \nonumber \\
   &|v| =   \sqrt{2\times ENERGY/ mass_p}.
\end{align}
Here $mass_p$ = 0.010438870 eV/$(\text{km/s})^2$ is the normalized proton mass and $v_{xyz}$ are the phase space velocities in the instrument (INST) frame and coordinate system. 
The velocity space integration weights corresponding to the VDF values are evaluated as 
\begin{equation}
\Delta^3 \mathbf{v} \ [(\text{km/s})^3 \text{ster}] = v^2  \Delta v \times \cos(THETA) \times \Delta THETA \times \Delta PHI
\end{equation}
where $v^2 \Delta v = \dfrac{\Delta ENERGY}{mass_p} \sqrt{\dfrac{2 \times ENERGY}{mass_p}}$ and $\Delta (THETA,PHI,ENERGY)$ are the corresponding bin widths of the SPANi L2 variables. 

Subsequently, $\mathbf{B}_{SPAN}$, the local mean magnetic field vector over the accumulation time period (TIME\_ACCUM variable in SPANi L2 data) around central time (TIME variable in SPANi L2 data) is evaluated for each SPANi measurement. 
The proton phase space velocities $v_{x,y,z}$ are transformed to a local mean \textbf{B}-aligned coordinate system, $v_{\parallel,\perp1,\perp2}$, using $\mathbf{B}_{SPAN}$. 
The proton temperature tensor is evaluated in the INST frame and then transformed to the plasma frame
\begin{align}
    \mathbf{T}_{ij,p} (INS&T) [\text{eV}] = \frac{mass_p \times \sum (v_iv_jf_p(\textbf{v})\Delta^3 \mathbf{v})}{n_p}, \ \ \ \text{where}  \ \ i,j = \parallel,\perp1,\perp2,  \nonumber  \\
    & \mathbf{T}_{ij,p} (plasma) [\text{eV}] = \mathbf{T}_{ij,p} (INST) - mass_p \times \mathbf{V}_p \mathbf{V}_p.
\end{align}
Here $n_p[\text{cm}^{-3}] = \sum(f_p(\mathbf{v}) \Delta^3\mathbf{v})$ is the density and $ \mathbf{V}_{i,p} [\text{km/s}]= \dfrac{\sum (v_i f_p(\mathbf{v}) \Delta^3\mathbf{v})}{n_p}$ are the mean velocity components. 
The temperature anisotropy is evaluated from the diagonal elements of $\mathbf{T}_{ij,p} (plasma)$ as 
$T_{\perp,p}/T_{\parallel,p}= \dfrac{T_{\perp1} + T_{\perp2}}{2 \times T_{\parallel}}$, where $T_{\perp1/2} $ are the temperatures along the two perpendicular directions to the local mean \textbf{B} and $T_{\parallel}$ is the temperature along the local mean \textbf{B}. 
We evaluate the parallel heat flux of protons in the INST frame and transform to the plasma frame as follows
\begin{align}
    q_{\parallel,p} (INST) [\text{eV } \text{km/s}] &= \dfrac{mass_p}{2} \times \frac{ \sum(v^2v_{\parallel} f_p(\mathbf{v})\Delta^3\mathbf{v})}{n_p}, \nonumber \\
    q_{\parallel,p} (plasma) [\text{eV } \text{km/s}]  = q_{\parallel,p} (INST) -    (\mathbf{V}_{i,p} &\cdot \mathbf{T}_{ij,p} (plasma))_{\parallel}  - \dfrac{1}{2} \text{Tr}(\mathbf{T}_{ij,p} (plasma)) V_{\parallel,p}  -  \dfrac{mass_p}{2}  V_p^2  V_{\parallel,p}.
\end{align}
The plasma frame heat flux values are then converted to SI units as $q_{\parallel,p} (plasma) [\text{kg } \text{(m/s)}^3 \text{cm}^{-3}] =  q_{\parallel,p} (plasma) [\text{eV km/s}] \times (m_p/mass_p)\times n_p \times 10^9$, where $m_p = 1.67\times 10^{-27}$ kg is the proton mass and  $10^9$ is the unit conversion factor of velocity from [km/s] to [m/s]. 
We normalize $q_{\parallel,p}$ with saturation heat flux \citep{Bale_2013_e_heat_flux, Halekas_2021_e_heat_flux},

\begin{equation}
    q_{0,p} [\text{kg} \text{(m/s)}^3 \text{cm}^{-3}] = \dfrac{3}{2}\ n_p T_p e v_{th,\parallel,p},
\end{equation}

where $T_p [eV] = \dfrac{\text{Tr}(\mathbf{T}_{ij,p}(plasma))}{3}$,  e = $1.602 \times 10^{-19}$ J/eV and $v_{th,\parallel,p} [\text{m/s}] = \sqrt{\dfrac{2 \times T_{p,\parallel}[\text{eV}]e}{m_p}}$.

\subsubsection{SPANi Field of View} \label{sec:meth_SPAN_FOV}
The SPANi proton moments are pre-screened for a good field of view (FOV), which is quantified by the parameter $A_{FOV}$ as described in \cite{Romeo_2024_thesis}, 
\begin{align}
    & A_{FOV} = A_\theta A_\phi , \ \ \ \  \text{where,} \ \ \ \
         A_x= CDF(z_{x_2}) - CDF(z_{x_1}), \nonumber  \\
         z_{x_1} = &\dfrac{x_1 - x_0}{w_x},  \ \ \ \ z_{x_2} = \dfrac{x_2 - x_0}{w_x},  \ \ \ \ 
         CDF(z) = \dfrac{1}{2} \left[1 + erf \left(\dfrac{z}{\sqrt{2}} \right) \right].  
\end{align}
Here, erf is the error function, and CDF is the cumulative distribution function. We evaluate best-fit Gaussian, $f(x) = h e^{-(x-x_0)^2/w_x^2}$ to EFLUX(x) where $x_0$ and $w_x$ are the mean and standard deviation, and $x_1$ and $x_2$ are FOV boundaries. 
We discard SPANi observations with $A_{FOV} < 0.75$ or $|\theta_{EFLUX\_max}|>40^\circ$ or $\phi_{EFLUX\_max} < 106^\circ$ or $\phi_{EFLUX\_max} > 164^\circ$, where $\theta_{EFLUX\_max}$ and $\phi_{EFLUX\_max}$ are the values of $\theta$ and $\phi$ where EFLUX($\theta/\phi$) is maximum, respectively. 
The SPANi-derived density, temperature anisotropy, and heat flux values are further screened by discarding observations that satisfy $n_{p,SPAN}/n_{e,QTN} < 0.85$, where $n_{e,QTN}$ is the quasi-thermal noise electron density. 
The solar wind velocities evaluated by SPANi are not screened, as they are less sensitive to FOV limitations.

\subsection{Quasi-thermal noise (QTN) Data} \label{sec:meth_QTN}
We use electron density from the PSP/FIELDS/RFS L3 \textit{sqtn\_rfs\_V1V2} data set \citep{Moncuquet_2020_PSP_FIELDS_QTN} for our analysis. 
We neglect the alpha particles, assume that the proton density ($n_p$) is equal to the QTN electron density ($n_{e,QTN}$), and use the latter to evaluate the parameters derived from proton density ( e.g., plasma beta, wave energy, instability threshold values).  
QTN data are unavailable for Encounters 8 and 11, and the analysis involving $n_{e,QTN}$ (Sections \ref{sec:F_wave_R_profile}, \ref{sec:F_wave_Vs_T_p_aniso}, and \ref{sec:F_wave_Vs_q_par}) excludes data from these encounters, except for the encounter-wise trend plots where SPANi $n_p$ are used when QTN data are unavailable.

\subsection{Anti-correlation Between Sampling Angle ($\theta_{BV}$) and Wave Observation} \label{sec:_meth_wave_theta_anti_corr}

\begin{figure*}
    \centering
    \includegraphics[width = 0.95\linewidth]{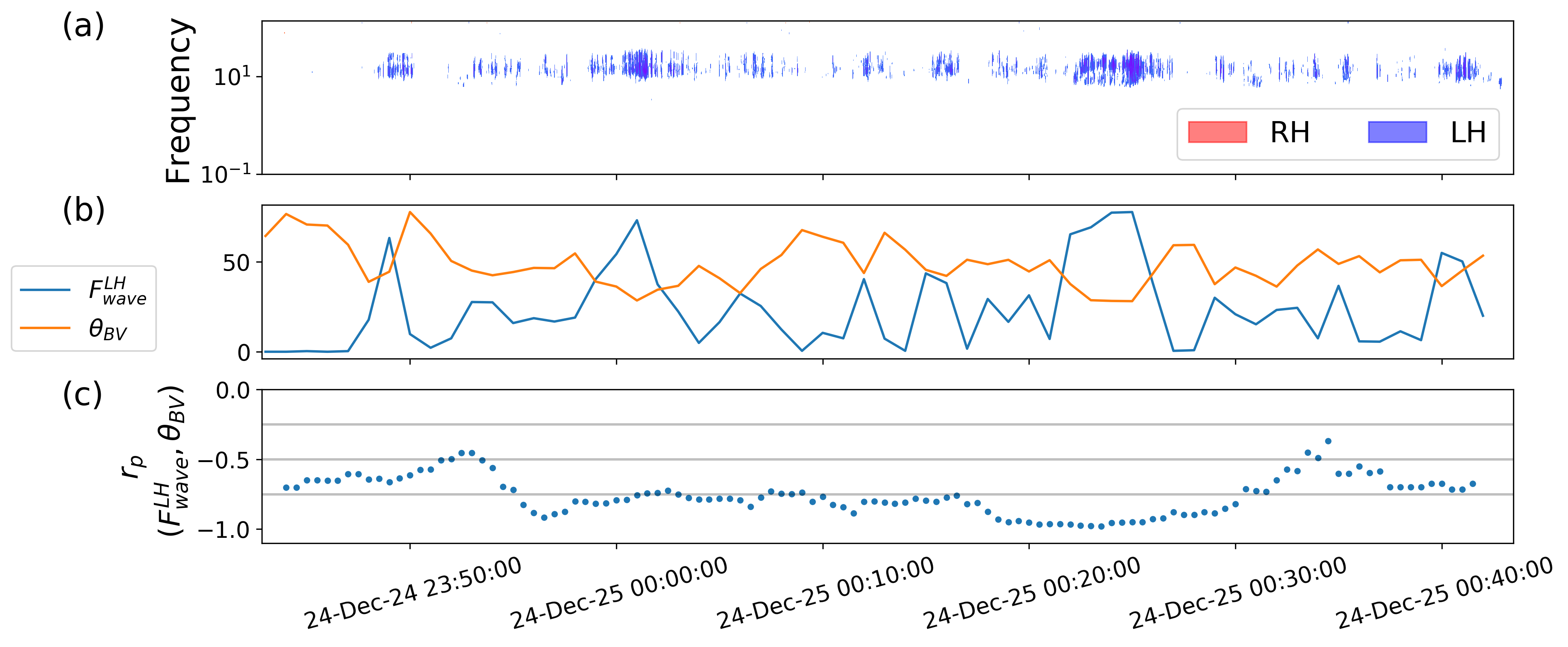}
    \caption{Sampling angle bias is evident in an hour-long observation of an LHW storm by PSP during Encounter 22.  
    (a) The smoothed circular polarization spectrum with $|\sigma| > 0.7$ as a function of frequency and time. 
    Plotted as a function of time are 
    (b) binned $F^{LH}_{wave}$(blue) and $\theta_{BV}$(orange) with a bin-width of 60 s, and 
    (c) Pearson correlation coefficient between $F^{LH}_{wave}$ and $\theta_{BV}$ over a sliding window of width 900 s. 
    The Pearson correlation coefficient consistently lies between -0.5 and -1. 
    $F^{LH}_{wave}$ curve (blue) in panel (b) is normalized by a factor such that its maximum value is equal to the maximum value of $\theta_{BV}$ (orange).}
    \label{fig:wave_theta_corr_example}

    \includegraphics[width = 0.95\linewidth]{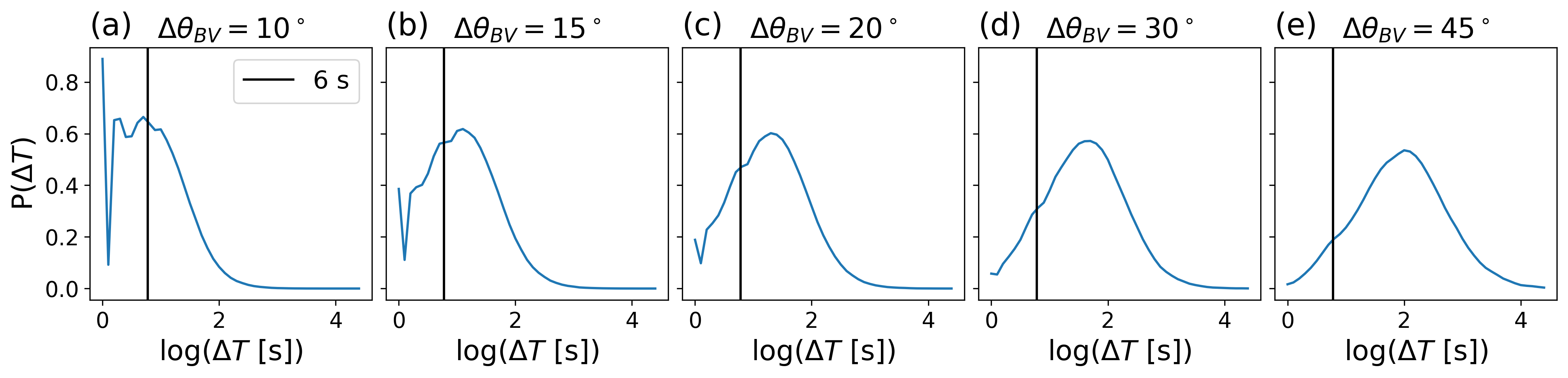}
    \caption{Probability distributions of $\Delta T$, the timescales over which the sampling angle, $\theta_{BV}$ rotates by an angle $\Delta \theta_{BV}$ for various values of $\Delta \theta_{BV}$.
    The black line indicates the $\sim 6$ second peak seen in the waiting time distribution shown in Fig.~\ref{fig:wave_waiting_residence_times}.}
    \label{fig:theta_BV_timescale_dist}
\end{figure*}

Whether ion-scale waves are observed/not observed depends on the sampling angle, i.e., the acute angle between solar wind velocity in the spacecraft frame and local magnetic field direction ($\theta_{BV}$). 
Waves are preferentially observed when $\theta_{BV}$ is small and may not be observed when $\theta_{BV}$ is large, even if they occur due to the omnipresent anisotropic turbulent fluctuations that have higher magnitudes when $\theta_{BV}$ approaches $90^\circ$ \citep{Bowen_2020_ICW}. 
To quantify this sampling effect, we evaluate the time fraction of wave observation, $F_{wave}^{(LH/RH)}$ (as described in Section \ref{sec:meth_PSP_FIELDS}) and binned $\theta_{BV}$ time series, both with a bin-width of 60 s. 
Sliding windows of width 900 s are considered over the binned time series, and the Pearson correlation coefficient ($r_p$) between $F_{wave}$ and $\theta_{BV}$ is evaluated within each sliding window if LHWs/RHWs are observed for at least 30 s within the window. 
The bin width of 60 s is chosen to represent the characteristic timescales of sampling angle variation. 
Varying the bin width or the sliding window width does not significantly change the resulting correlation. 
Figure \ref{fig:wave_theta_corr_example} shows the strong sampling angle bias for an example 1-hour interval from Encounter 22 when a LHW storm is observed, where the magnitude of $r_p$ is consistently large and negative. 
We infer that the LHW is observed (not observed) by PSP when the sampling angle is smaller (larger), suggesting that the observed wave possibly occurs throughout the 1-hour interval but is not seen at all times.

Furthermore, we evaluate $P(\Delta T)$, the distribution of timescales for the sampling angle $\theta_{BV}$ to fluctuate over various values. 
SPANi observations at perihelia are sampled at a set of time increments, [1.75, 3.5, 7] s. 
The evaluation of $P(\Delta T)$ at $\Delta T$ near these SPANi cadences is limited by the time resolution. 
To overcome this limitation, we resample the SPAN-derived $\theta_{BV}$ time series at a cadence of 0.5 s using cubic interpolation. Starting from the first time point, $\Delta T$ are the times taken for $\theta_{BV}$ to increase or decrease by a specified angle, $\Delta \theta_{BV}$. We estimate the distribution of $\Delta T$ using the kernel density estimation method with Gaussian kernels and the ``Scott's'' bandwidth for a set of $\Delta \theta_{BV}$ values as shown in Figure \ref{fig:theta_BV_timescale_dist}. The black line corresponds to the time scale at which LHW waiting and residence times are maximum, as discussed in Section \ref{sec:waiting_residence_times}.

\subsection{Kinetic Microinstabilities}

Instabilities are mechanisms that transfer free energy from non-thermal structures to plasma normal modes, which scatter plasma particles and reduce the free energy. 
Kinetic microinstabilities \citep{Gary_1993_plasma_microinstabilities_book, Verscharen_2019_Review} are driven by non-Maxwellian structures of particle VDFs, including anisotropies in temperatures with respect to local \textbf{B} direction and the presence of drifting subpopulations. 
Kinetic microinstabilities are mediated by the interaction between waves and a fraction of resonant particles whose thermal velocities are similar to the wave’s phase velocity. 
Several kinetic instabilities drive ion-scale circularly polarized waves (ICWs and FMWs). 
Previous works have developed parametric models for instability thresholds as a function of parameters that quantify the deviation from LTE using analytical and computational methods (using linear plasma dispersion solvers) \citep{Gary_1994_instability_model, Hellinger_2006_T_aniso_instability_thresh, Verscharen_2016_instability_thresh}. 
In this work, we focus on proton-driven instabilities.

\subsubsection{Temperature Anisotropy Instabilities} \label{sec:meth_Tp_aniso_inst}
The proton temperature anisotropy instabilities constrain the deviation of proton temperature anisotropy ($T_{\perp,p}/T_{\parallel,p}$) from unity. 
The ion  cyclotron instability (parallel firehose instability) is active when $T_{\perp,p}/T_{\parallel,p} > 1 \ ( < 1)$ and generates ICWs (FMWs). 
In this work, we statistically analyze $F^{(LH/RH)}_{wave}$ with respect to these temperature anisotropy instability thresholds, which take the form of

\begin{equation} \label{eq:T_p_aniso_thresh}
    \frac{T_{\perp,p}}{T_{\parallel,p}}  = 1 + \frac{a}{(\beta_{\parallel,p}-c)^b}
\end{equation}
where $\beta_{\parallel,p}$ is the ratio of the thermal pressure of the protons along the local \textbf{B} direction to the magnetic pressure, and the parameters a,b, and c are given in Table 1 of \cite{Verscharen_2016_instability_thresh} for $\gamma/\Omega_p = 10^{-3}$ where $\gamma$ is the wave growth rate and $\Omega_p$ is the proton gyrofrequency. 
The mirror and oblique firehose instability thresholds are included as well, although they do not generate ion-scale circularly polarized waves.

\subsubsection{Beam Instabilities} \label{sec:meth_SAVIC}

The presence of ion beams can drive plasma instabilities \citep{Gary_1993_plasma_microinstabilities_book, Verscharen_2013_Oblique_modes_for_beams}. 
Proton beams can drive both ICWs and FMWs unstable \citep{Daughton_Gary_p_p_instabilities, Pezzini_2024_proton_beam_instability_simulation}. 
However, at low $\beta_{p,\parallel}$, which is typically observed at the PSP perihelia, we expect proton beams to mainly drive parallel and oblique FMWs unstable \citep{ Verniero_2020_waves_proton_beams, Verniero_2022_hammerhead, Martinović_2025}. 
Assuming bi-Maxwellian VDFs, the stability of FMWs depends on a set of parameters, $\mathcal{P} = \left[\beta_{\parallel,pc}, \dfrac{T_{\perp,pc}}{T_{\parallel,pc}}, \dfrac{T_{\parallel,pc}}{T_{\parallel,pb}}, \dfrac{T_{\perp,pb}}{T_{\parallel,pb}}, \dfrac{n_{pb}}{n_{pc}}, \dfrac{v_{bc}}{v_A} \right]$. 
Here, the subscripts pc and pb refer to proton core and beam, respectively, $v_{bc}$ is the relative drift velocity of the beam in the core reference frame and $v_A$ is the local Alfv\'en speed. 
In this work, we reduce this multi-dimensional parameter space to a single parameter, i.e., the normalized heat flux (see Section \ref{sec:meth_PSP_SPAN} and \ref{sec:q_par_bi_Maxwellian}). 
Using observations, we infer a threshold value of normalized heat flux above which FMWs are likely to become unstable. 
We further validate the inferred instability threshold value by using the \texttt{SAVIC} machine learning algorithm that predicts stability of the multi-dimensional parameter space associated with plasma VDFs.

%subsection{\texttt{SAVIC}} 

Stability Analysis Vitalizing Instability Classification (\texttt{SAVIC}, \cite{Martinović_2023_SAVIC}) is a machine learning algorithm that predicts the type of plasma microinstability by recognizing the maximum unstable plasma mode and the VDF component that drives this mode for a given set of plasma parameters. 
\texttt{SAVIC} is trained on calculations from a combination of \texttt{PLUME} and \texttt{PLUMAGE} solvers. The \texttt{PLUME} linear plasma dispersion solver \citep{Klein_Howes_2015_PLUME, Klein_2025_PLUME} determines the dispersion solution for plasma modes in a hot, collisionless plasma, and the \texttt{PLUMAGE} solver \citep{Klein_2017_Nyquist} evaluates the maximum unstable mode for a given set of plasma parameters. 
\texttt{SAVIC} reports whether a set of VDFs is stable or unstable, determines the kind of instability (e.g. a FMW or ICW) as well as the growth rate and other characteristics of the unstable wave.

\section{Heat flux of proton core and beam  bi-Maxwellian distribution} \label{sec:q_par_bi_Maxwellian}

In this section, we derive the expression for the normalized heat flux of a two-component proton core and beam bi-Maxwellian distribution. We consider $\Sigma_i n_i \mathcal{M}_i(U_i,v_{i,th,\perp},v_{i,th,\parallel})$, a mixture of multiple bi-Maxwellian velocity distribution functions that drift relative to each other in a direction parallel to the local magnetic field, where $n_i$, $U_i$, and $m_i$ are the densities, average parallel velocities, and masses of the \textit{i}th species, respectively. Let $n = \sum n_i$ be the total density of the mixture.
The mean parallel velocity of the mixture is given by 
\begin{equation} \label{eq:mixture_dens}
    U = \dfrac{\Sigma_i n_i U_i}{\Sigma_in_i} = \dfrac{\Sigma_i n_i U_i}{n}.
\end{equation}
The random velocity of an \textit{i}th species particle with velocity $v_i$ in the \textit{i}th species center of mass (COM) frame is
$c_i = v_i - U_i$.
The random velocities of a particle of the \textit{i}th species in the mixture COM frame (relative to U) are transformed to the \textit{i}th species COM frames as 
\begin{equation} \label{eq:v_mixture_frame_to_ith_frame}
    v_i - U =  v_i - U_i + \Delta U_i, \ \ \ \ \Delta U_i = U_i - U.
\end{equation}

\subsection{Parallel and Perpendicular Temperatures of the Mixture}
The parallel and perpendicular temperatures of the \textit{i}th bi-Maxwellian species in their respective COM frames are
\begin{align} \label{eq:ith_Tpar}
 k_B T_{\parallel,i} & =   m_i v^2_{i,th,\parallel}   = m_i <{c}^2_{\parallel, i}>,  \\ \label{eq:ith_Tperp}
 k_B T_{\perp, i} = \dfrac{k_B T_{\perp1, i} + k_B T_{\perp2, i}}{2} &   = m_i \dfrac{ <{c}^2_{\perp1, i}> + <{c}^2_{\perp2, i}>}{2} = \dfrac{m_i <{c}^2_{\perp, i}>}{2}, 
\end{align}
where $<>$ denotes the expectation value. Note that the average perpendicular thermal velocities include both perpendicular velocity components.
For a bi-Maxwellian mixture with only parallel drifts, the perpendicular temperature in the mixture COM frame is 
$T_\perp = \dfrac{1}{n} \sum_i  n_i T_{\perp,i}$.
For a proton core and beam mixture, the above equation is written as
\begin{align}\label{eq:T_perp_bc}
    T_\perp =  \dfrac{n_c T_{\perp,c} + n_b T_{\perp,b}}{n_c + n_b}.
\end{align} 
The parallel temperature of a bi-Maxwellian mixture in the mixture COM frame is
$$k_BT_\parallel= \dfrac{1}{n} \sum_i n_i m_i <(v_{\parallel, i} - U)^2>.$$
Substituting for $v_{\parallel, i} - U$ using Equation \ref{eq:v_mixture_frame_to_ith_frame}, equating odd moments of bi-Maxwellian to 0, and using Equation \ref{eq:ith_Tpar}
\begin{align}
    k_BT_\parallel= \dfrac{1}{n} \sum_i n_i  \left( k_B T_{\parallel,i} + m_i \Delta U_i^2  \right) .\nonumber
\end{align}
For a proton core and beam mixture, the above equation is written as follows
\begin{align}\label{eq:T_par_cb_init}
    k_B T_\parallel= \dfrac{k_B}{n_c + n_b} ( n_c T_{\parallel,c} + n_b T_{\parallel, b})  + \dfrac{m_p}{n_c + n_b} \left(n_b (U_b - U)^2 + n_c (U_c - U)^2 \right). 
\end{align}
Using Equation \ref{eq:mixture_dens} the proton core and beam velocities in the mixture COM frame are expressed in terms of beam drift velocity relative to the core
\begin{align}
    U_b - U = U_b - \dfrac{n_cU_c + n_bU_b}{n_c + n_b} = \dfrac{n_c (U_b - U_c)}{n_c + n_b} = \dfrac{n_c \Delta U_{bc}}{n_c + n_b} ,\label{eq:ub-u} \\ 
    U_c - U = U_c - \dfrac{n_cU_c + n_bU_b}{n_c + n_b} = \dfrac{-n_b(U_b - U_c)}{n_c+n_b} = \dfrac{-n_b \Delta U_{bc}}{n_c+n_b} .\label{eq:uc-u}
\end{align}
Substituting Equations \ref{eq:ub-u} and \ref{eq:uc-u} into Equation \ref{eq:T_par_cb_init}, the parallel temperature of the proton core and beam distribution is written as
\begin{align}\label{eq:T_par_bc}
    k_B T_\parallel= \dfrac{k_B}{n_c + n_b} ( n_c T_{\parallel,c} + n_b T_{\parallel, b}) + \dfrac{m_p}{(n_c + n_b)^2} \left( n_b n_c \Delta U^2_{bc} \right).
\end{align}

\subsection{Parallel Heat Flux of the Mixture}

The total parallel heat flux is the sum of the parallel heat fluxes of each species with respect to the mixture COM frame, $q_\parallel = \sum_i q_{\parallel, i} $. The heat flux of the \textit{i}th species in the mixture COM frame is
\begin{align}
q_{\parallel, i} = \dfrac{1}{2}m_i n_i \left< \left( (v_{\parallel,i} - U)^2 + v^2_{\perp, i} \right)(v_{\parallel,i} - U) \right>.
\end{align}
Substituting for $v_i - U$ in the above equation using Equation \ref{eq:v_mixture_frame_to_ith_frame}, equating odd moments to 0, and using Equations \ref{eq:ith_Tpar} and \ref{eq:ith_Tperp}, the total heat flux is
\begin{align}
    q_{\parallel} = \sum_i \dfrac{n_i}{2} \left [  3 \Delta U_i k_B T_{\parallel, i} + m_i \Delta U_i^3
+ \Delta U_i 2 k_B T_{\perp,i} \right]   .
\end{align}
For a proton core and beam mixture, the above equation takes the form
\begin{align}
    q_{\parallel} = \dfrac{3 k_B}{2} \left[ n_c(U_c - U) T_{\parallel, c} + n_b (U_b - U) T_{\parallel, b} \right] + &\dfrac{m_p}{2}\left[ n_c (U_c-U)^3 + n_b (U_b-U)^3   \right] + \\ &k_B \left[ n_c(U_c - U) T_{\perp, c} + n_b (U_b - U) T_{\perp, b} \right] .\nonumber
\end{align}
Using Equations \ref{eq:ub-u} and \ref{eq:uc-u}, the parallel heat flux of the proton core and beam mixture is written as

\begin{align}
    q_{\parallel} = \dfrac{3 k_B n_c n_b \Delta U_{bc}}{2 (n_c + n_b)} \left( T_{\parallel, b} - T_{\parallel, c} \right) + \dfrac{m_p n_b n_c \Delta U_{bc}^3 \left( n_c^2 - n_b^2 \right)}{2 (n_c + n_b)^3} + \dfrac{k_B n_c n_b \Delta U_{bc}}{n_c + n_b} \left( T_{\perp, b} - T_{\perp, c} \right).
\end{align}

\subsection{Normalization}

The expressions of the parallel and perpendicular temperatures and heat flux of the proton core and beam mixture derived above are written in terms of the following dimensionless parameters used in PLUME
$$\beta_{\parallel,c} = \dfrac{v_{th,\parallel,c}}{v_{A,c}}, \ \ \ \ \aleph_c = \dfrac{T_{\perp, c}}{T_{\parallel, c}}, \ \ \ \ \aleph_b = \dfrac{T_{\perp, b}}{T_{\parallel, b}}, \ \ \ \ v_b = \dfrac{U_b - U_c}{v_{A,c}}, \ \ \ \ D_b = \dfrac{n_b}{n_c}, \ \ \ \ \tau_b = \dfrac{T_{\parallel, c}}{T_{\parallel, b}}.$$
The expression for $T_\perp$ in Equation \ref{eq:T_perp_bc} is written in terms of dimensionless parameters as 
\begin{align} \label{eq:T_perp_dimensionless}
    \dfrac{T_\perp}{T_{\parallel, c}} = \dfrac{ \aleph_c + \dfrac{D_b\aleph_b}{\tau_b} }{1 + D_b}.
\end{align}
Similarly, the expression for $T_\parallel$ in Equation \ref{eq:T_par_bc}  is written as
\begin{align} \label{eq:T_par_dimensionless}
    \dfrac{T_\parallel}{T_{\parallel, c}} = \left[ \dfrac{1 + \dfrac{D_b}{\tau_b}}{1+D_b} + \dfrac{ D_b v_b^2}{(1+D_b)^2} \dfrac{2}{\beta_{\parallel,c}}\right].
\end{align}
The parallel heat flux is written in terms of dimensionless parameters as
\begin{align} \label{eq:q_par_dimensionless}
    q_{\parallel} = n_c k_B T_{\parallel, c}v_{A,c}  \left[ \dfrac{3    v_b}{2 \left(\dfrac{1}{D_b} + 1 \right)} \left( \dfrac{1}{\tau_b}  - 1 \right) + \dfrac{v_b^3 \left( \dfrac{1}{D_b}  - 1 \right) }{\left(\dfrac{1}{D_b} + 1 \right)^2} \dfrac{1}{\beta_{\parallel, c}} + \dfrac{  v_b}{\dfrac{1}{D_b} + 1} \left(  \dfrac{\aleph_b}{\tau_b} - \aleph_c \right) \right].
\end{align}
Furthermore, the parallel heat flux is normalized with the saturation heat flux
\begin{align}
    q_0 = \dfrac{3}{2}nk_B T v_{th, \parallel} = \dfrac{3}{2} (n_c + n_b) \dfrac{k_B(T_\parallel + 2T_\perp)}{3}\sqrt{\dfrac{2k_BT_\parallel}{m_p}}. \nonumber
\end{align}
The saturation heat flux is expressed in terms of dimensionless parameters as 
\begin{align} \label{eq:saturation_q_par_dimensionless}
    q_0 = \dfrac{1}{2} n_c k_B T_{\parallel, c} \sqrt{\dfrac{2k_BT_{\parallel, c}}{m_p}}  (1 + D_b) \dfrac{T_\parallel + 2T_\perp}{T_{\parallel, c}}\sqrt{ \dfrac{T_\parallel}{T_{\parallel, c}}}.
\end{align}
Thus, using Equations \ref{eq:q_par_dimensionless} and \ref{eq:saturation_q_par_dimensionless}, the normalized heat flux is expressed as
\begin{align}
    \dfrac{q_\parallel}{q_0} =  \dfrac{2}{\sqrt{\beta_{\parallel,c}}}
    \dfrac{\left[ \dfrac{3    v_b}{2 \left(\dfrac{1}{D_b} + 1 \right)} \left( \dfrac{1}{\tau_b}  - 1 \right) + \dfrac{v_b^3 \left( \dfrac{1}{D_b}  - 1 \right) }{\left(\dfrac{1}{D_b} + 1 \right)^2} \dfrac{1}{\beta_{\parallel, c}} + \dfrac{  v_b}{\dfrac{1}{D_b} + 1} \left(  \dfrac{\aleph_b}{\tau_b} - \aleph_c \right) \right]}{(1 + D_b) \dfrac{T_\parallel + 2T_\perp}{T_{\parallel, c}}\sqrt{ \dfrac{T_\parallel}{T_{\parallel, c}}} },
\end{align}
where $\dfrac{T_\parallel}{T_{\parallel, c}}$ and $\dfrac{T_\perp}{T_{\parallel, c}}$ are given by the Equations \ref{eq:T_par_dimensionless} and \ref{eq:T_perp_dimensionless}, respectively.

%\input{Appendix}

%Co-authors: comment out the appendix for faster compilation.

\section{Encounter-wise trends for all other Encounters}
\label{sec:Ewise}

In this subsection, we extend the Encounter 12 presentation in Figure~\ref{fig:E12_encounterwise_trends} to encounter-wise trends in the same format for Encounters 3 - 24. 
As QTN observations are unavailable for Encounters 8 and 11, the corresponding encounter-wise plots are evaluated by using proton density as evaluated by SPANi observations. 

\begin{figure}
    \centering
    \includegraphics[width=0.95\linewidth]{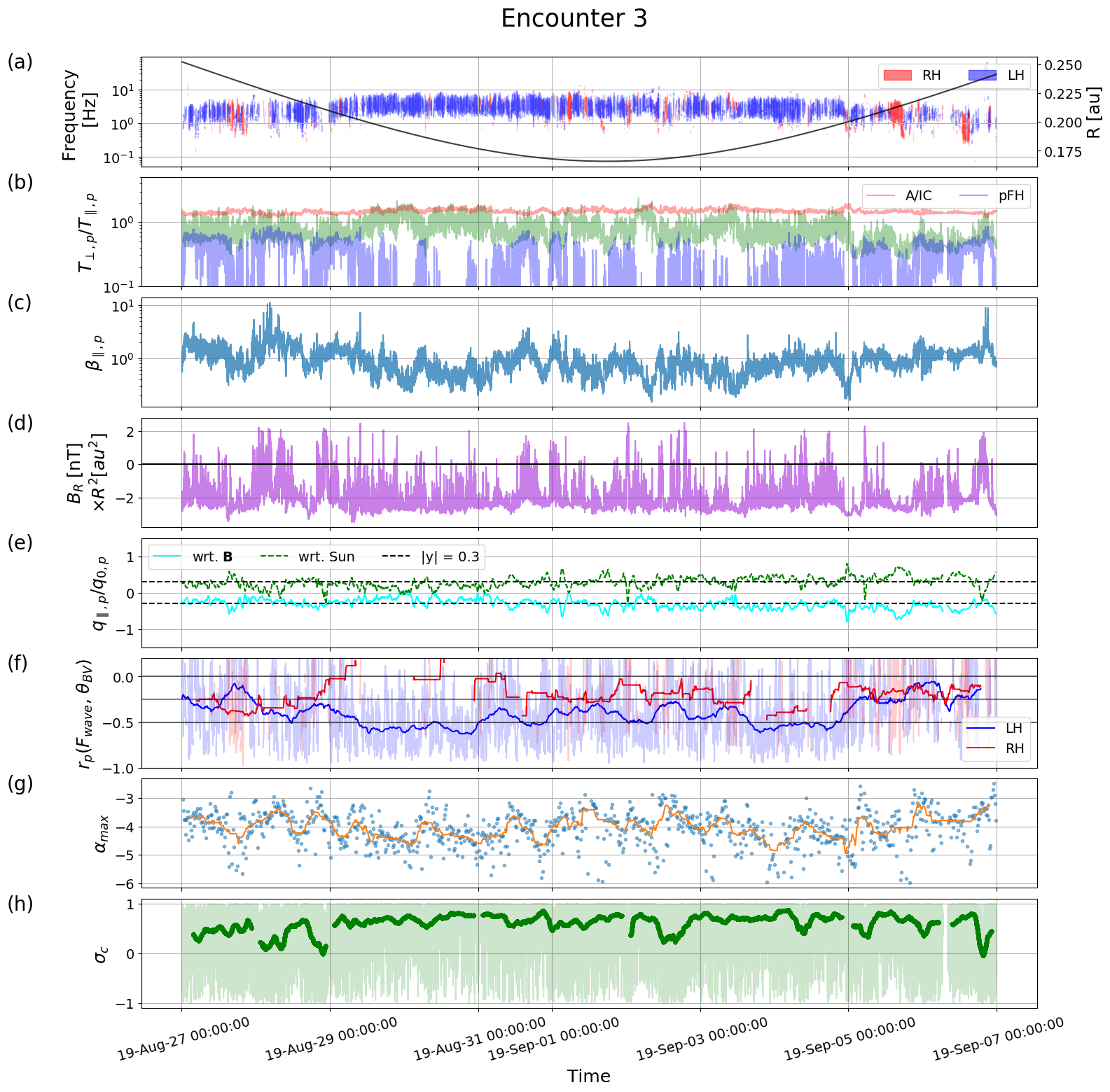}
    \caption{Trends for Encounter 3, organized as Figure \ref{fig:E12_encounterwise_trends}}
    \label{fig:App_Ewise_E3}
\end{figure}

\begin{figure}
    \centering
    \includegraphics[width=0.95\linewidth]{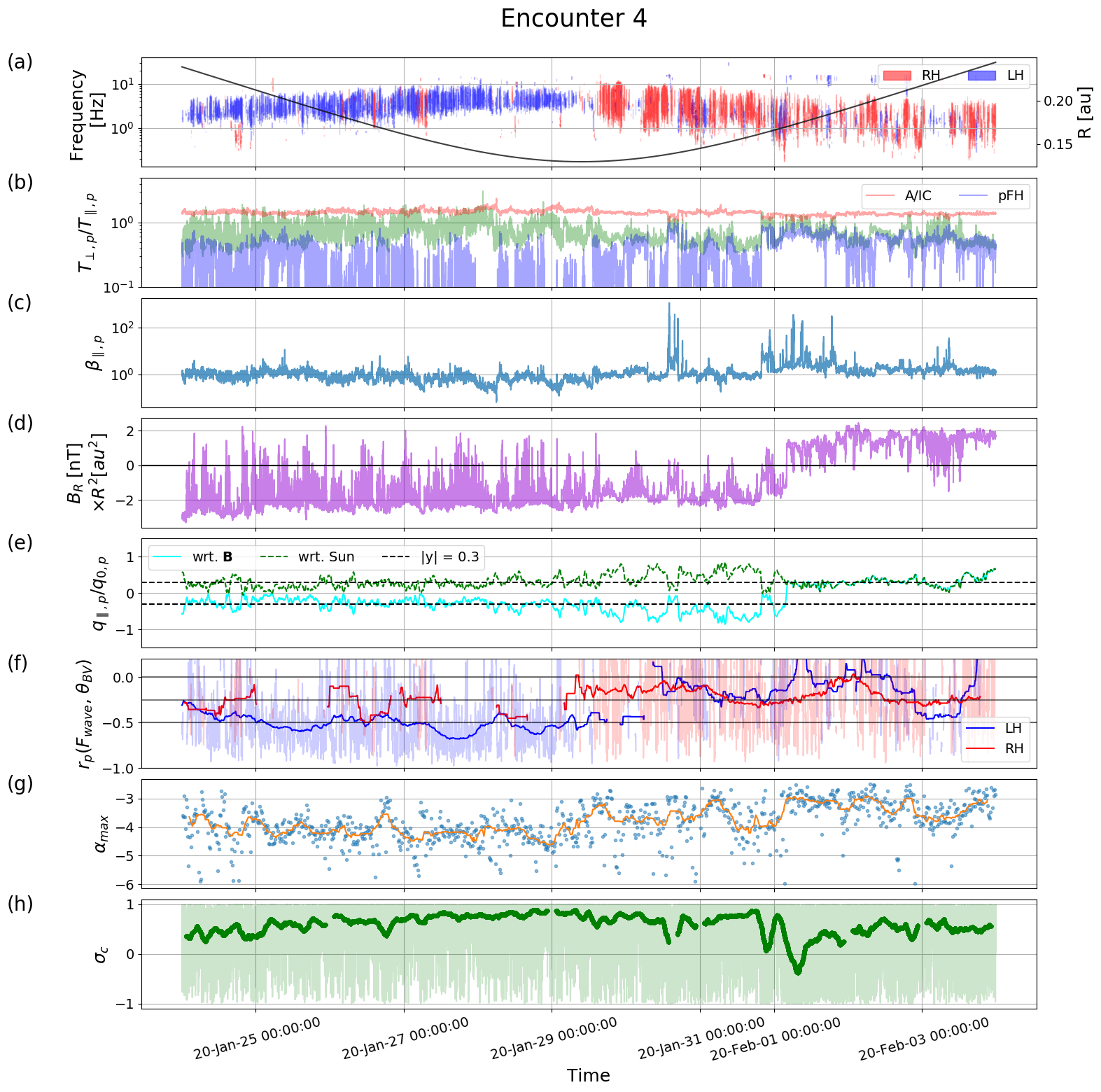}
    \caption{Trends for Encounter 4, organized as Figure \ref{fig:E12_encounterwise_trends}}
    \label{fig:App_Ewise_E4}
\end{figure}

\begin{figure}
    \centering
    \includegraphics[width=0.95\linewidth]{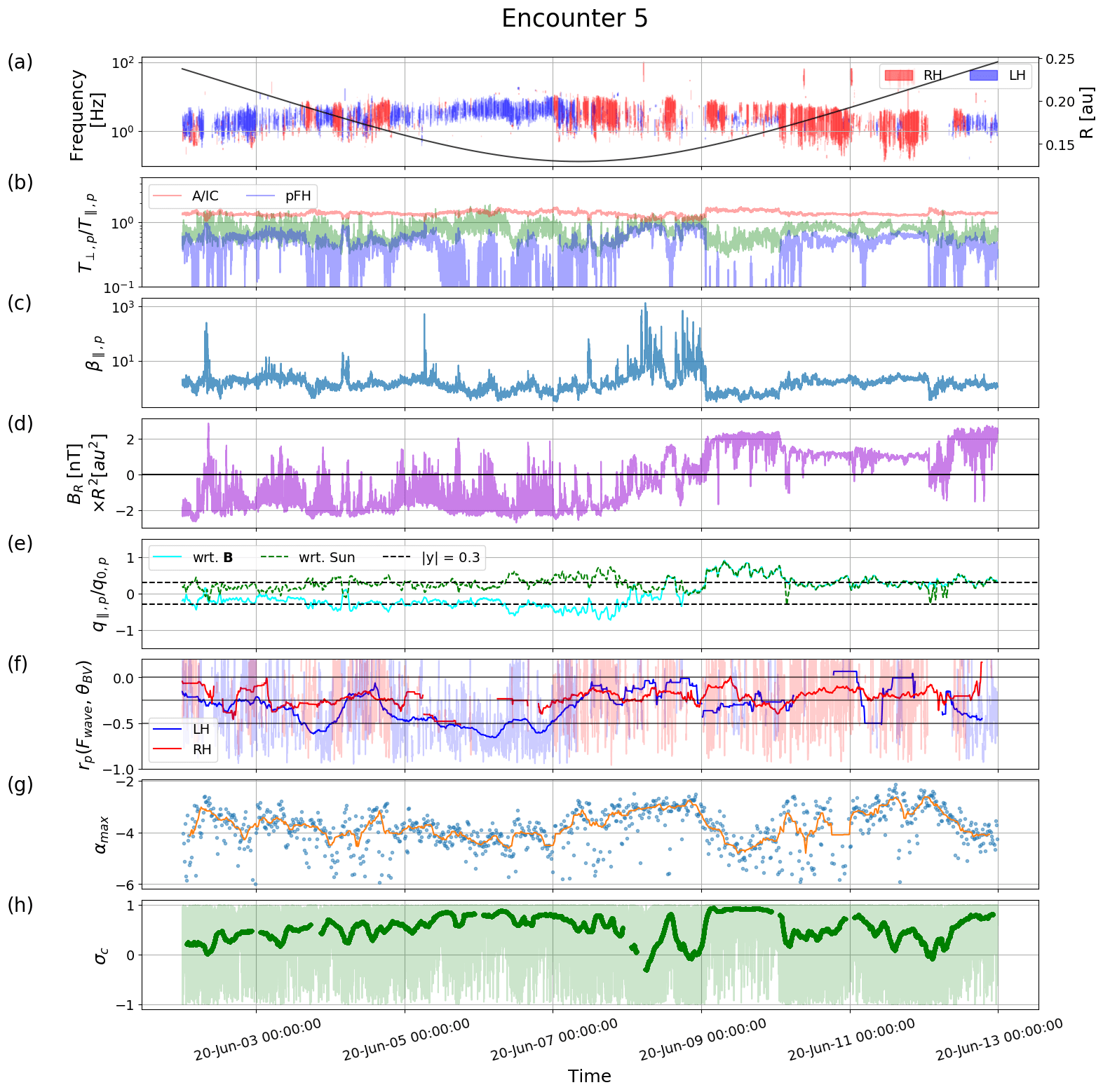}
    \caption{Trends for Encounter 5, organized as Figure \ref{fig:E12_encounterwise_trends}}
    \label{fig:App_Ewise_E5}
\end{figure}

\begin{figure}
    \centering
    \includegraphics[width=0.95\linewidth]{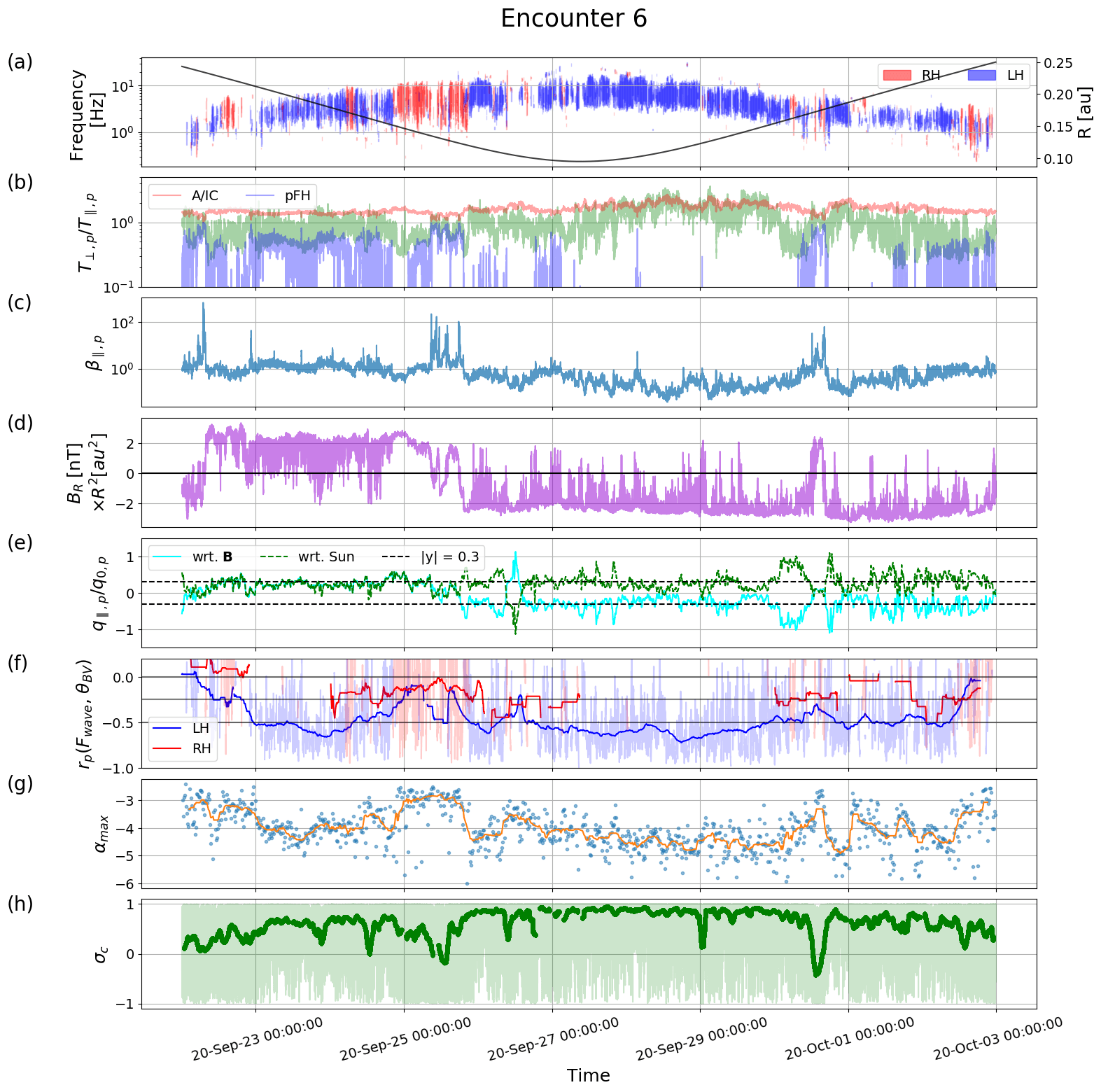}
    \caption{Trends for Encounter 6, organized as Figure \ref{fig:E12_encounterwise_trends}}
    \label{fig:App_Ewise_E6}
\end{figure}

\begin{figure}
    \centering
    \includegraphics[width=0.95\linewidth]{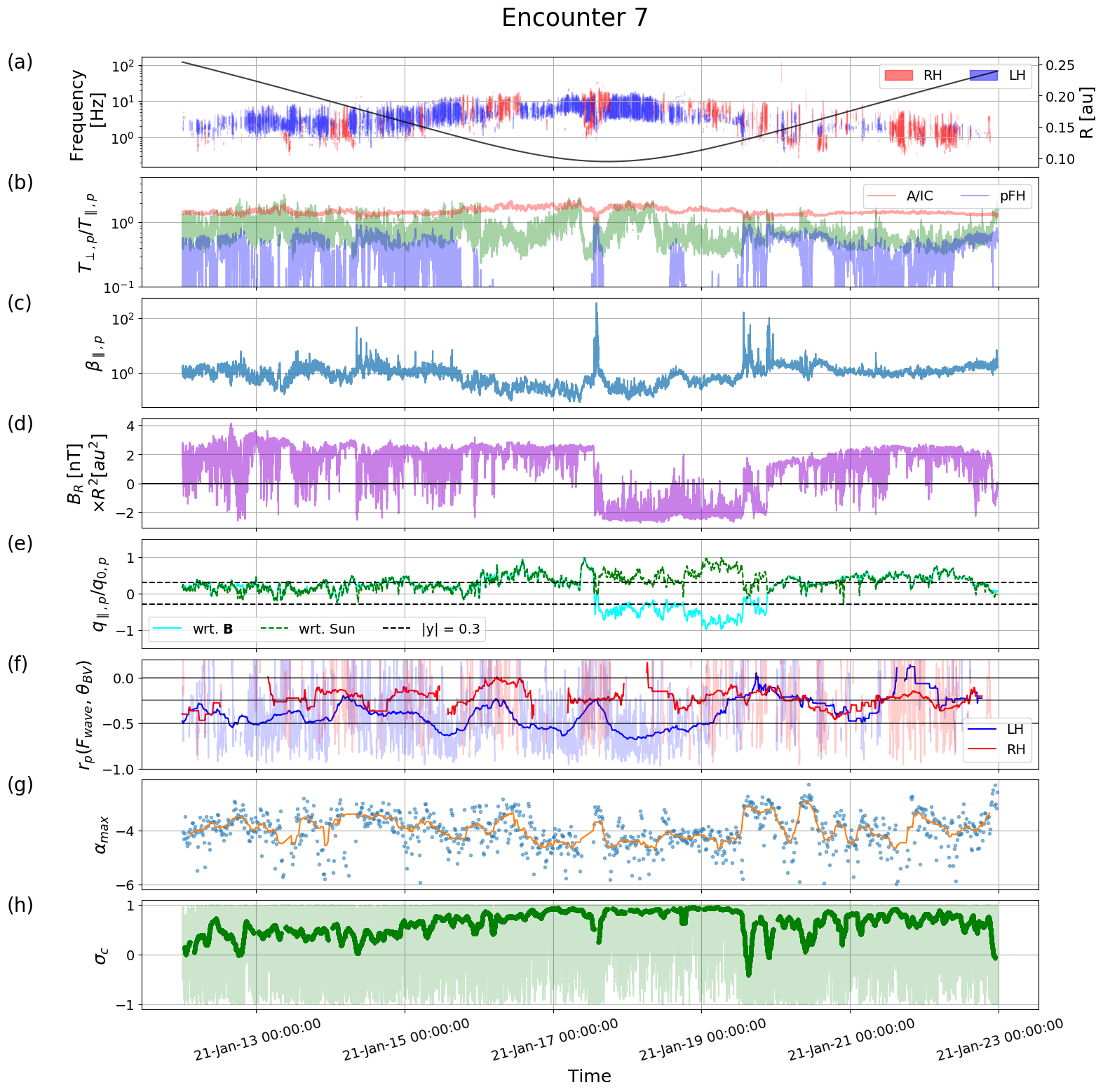}
    \caption{Trends for Encounter 7, organized as Figure \ref{fig:E12_encounterwise_trends}}
    \label{fig:App_Ewise_E7}
\end{figure}

\begin{figure}
    \centering
    \includegraphics[width=0.95\linewidth]{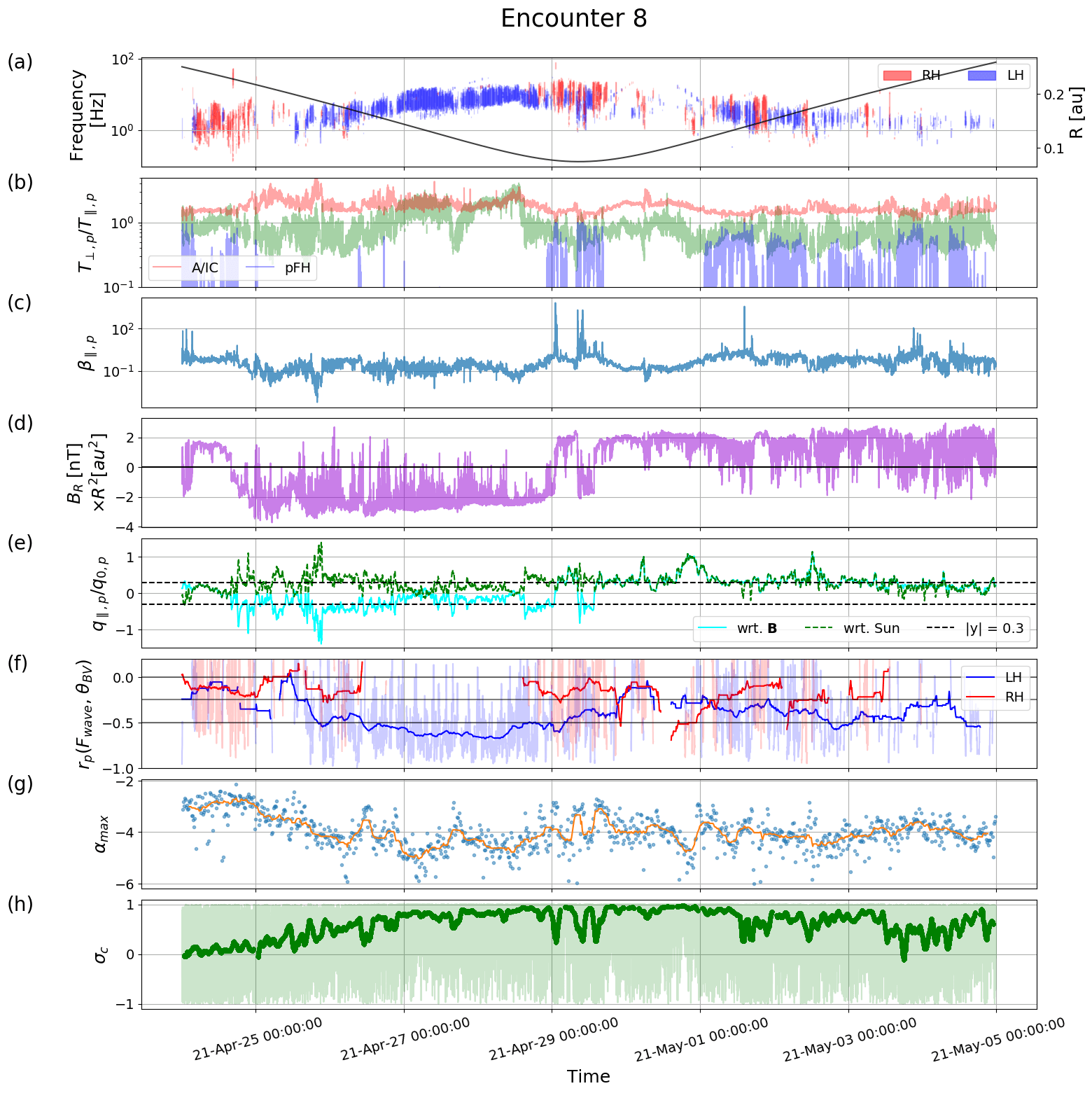}
    \caption{Trends for Encounter 8, organized as Figure \ref{fig:E12_encounterwise_trends}}
    \label{fig:App_Ewise_E8}
\end{figure}

\begin{figure}
    \centering
    \includegraphics[width=0.95\linewidth]{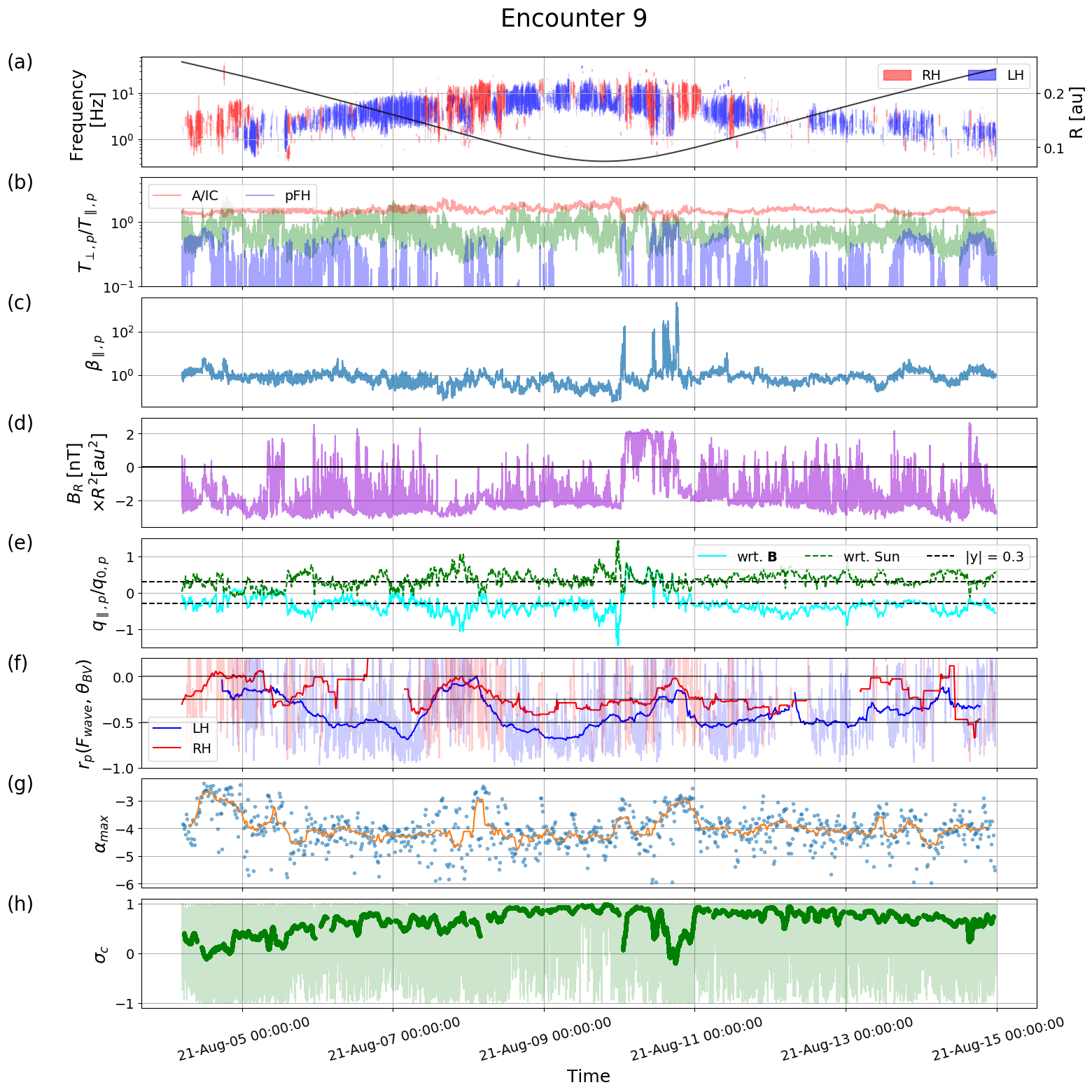}
    \caption{Trends for Encounter 9, organized as Figure \ref{fig:E12_encounterwise_trends}}
    \label{fig:App_Ewise_E9}
\end{figure}

\begin{figure}
    \centering
    \includegraphics[width=0.95\linewidth]{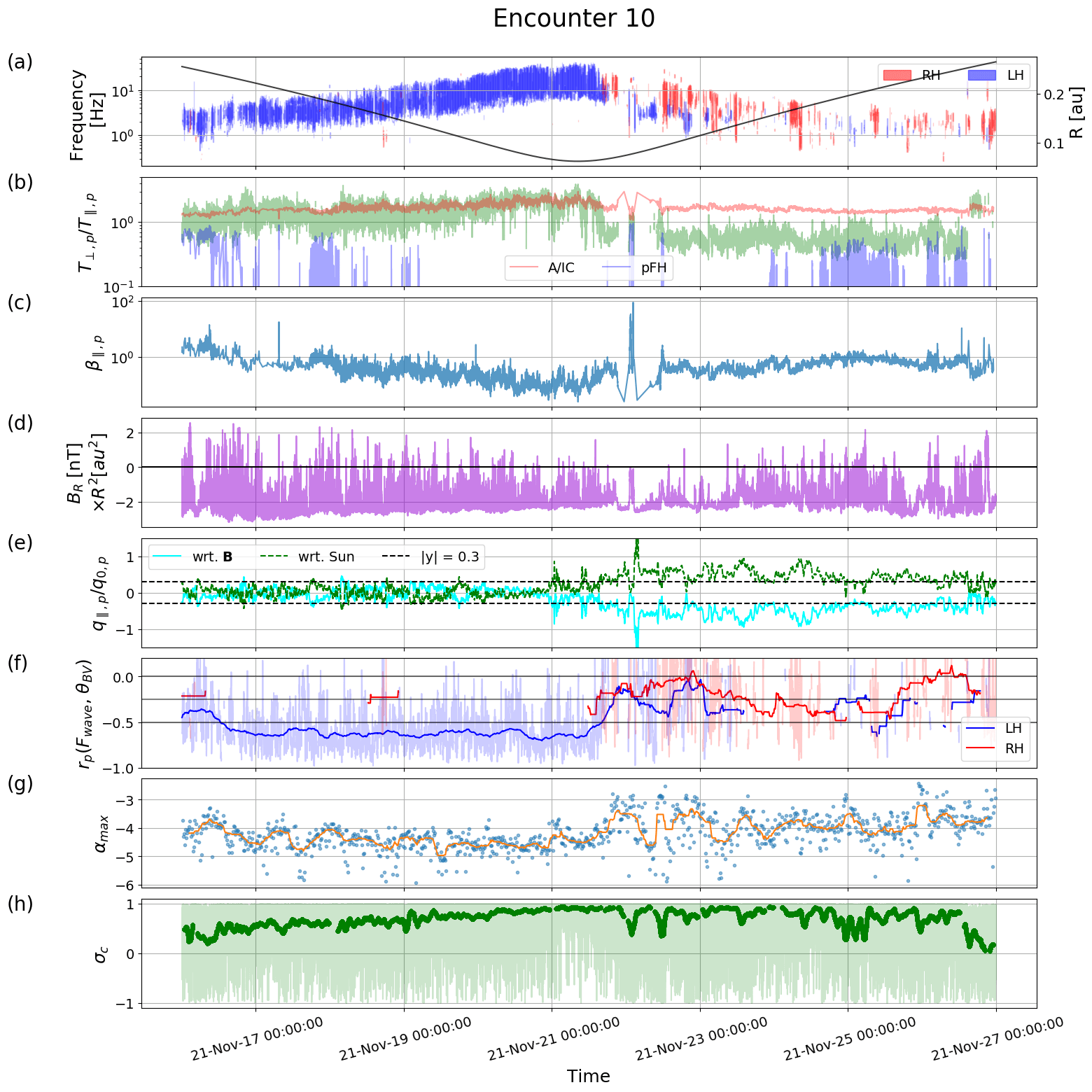}
    \caption{Trends for Encounter 10, organized as Figure \ref{fig:E12_encounterwise_trends}}
    \label{fig:App_Ewise_E10}
\end{figure}

\begin{figure}
    \centering
    \includegraphics[width=0.95\linewidth]{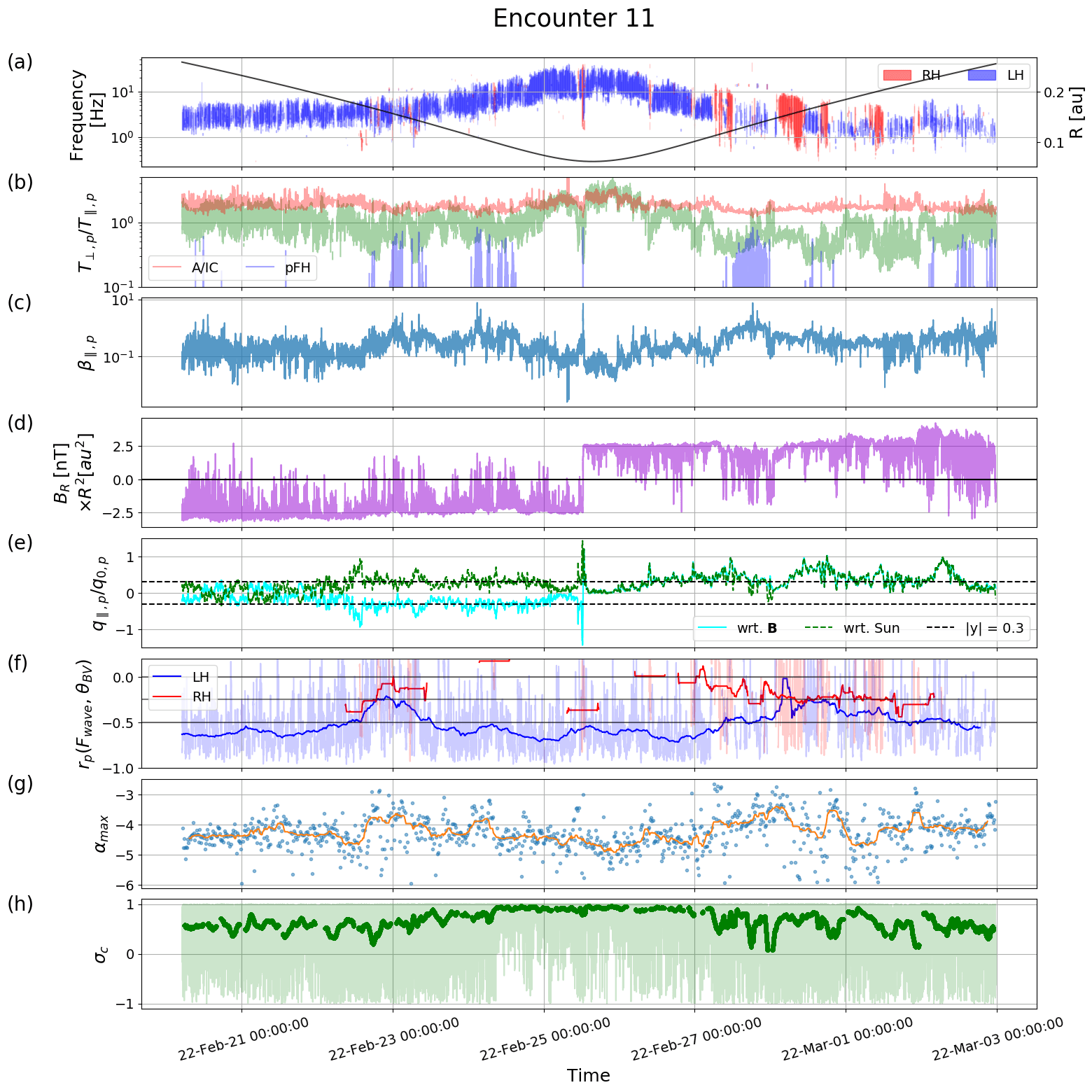}
    \caption{Trends for Encounter 11, organized as Figure \ref{fig:E12_encounterwise_trends}}
    \label{fig:App_Ewise_E11}
\end{figure}

\begin{figure}
    \centering
    \includegraphics[width=0.95\linewidth]{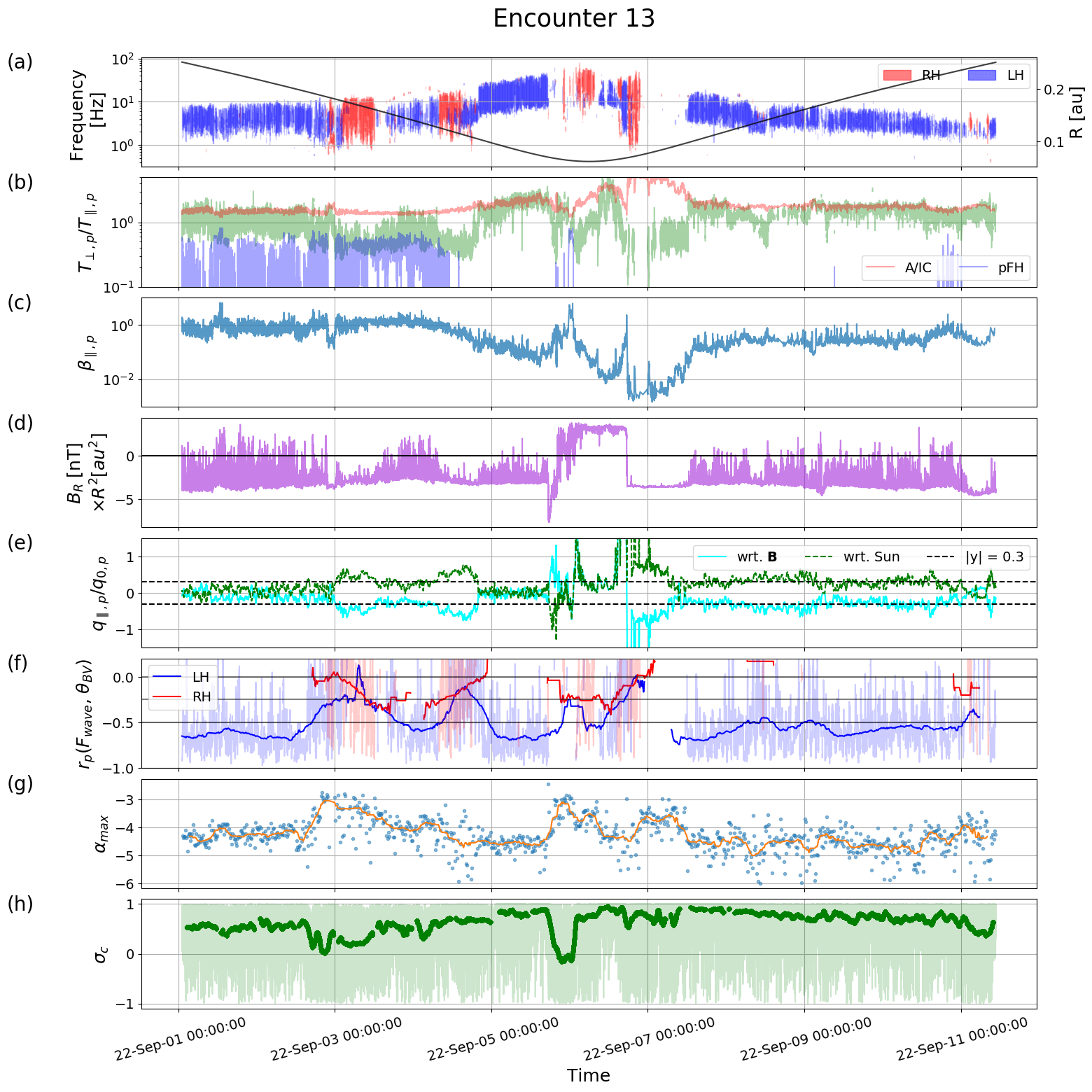}
    \caption{Trends for Encounter 13, organized as Figure \ref{fig:E12_encounterwise_trends}}
    \label{fig:App_Ewise_E13}
\end{figure}

\begin{figure}
    \centering
    \includegraphics[width=0.95\linewidth]{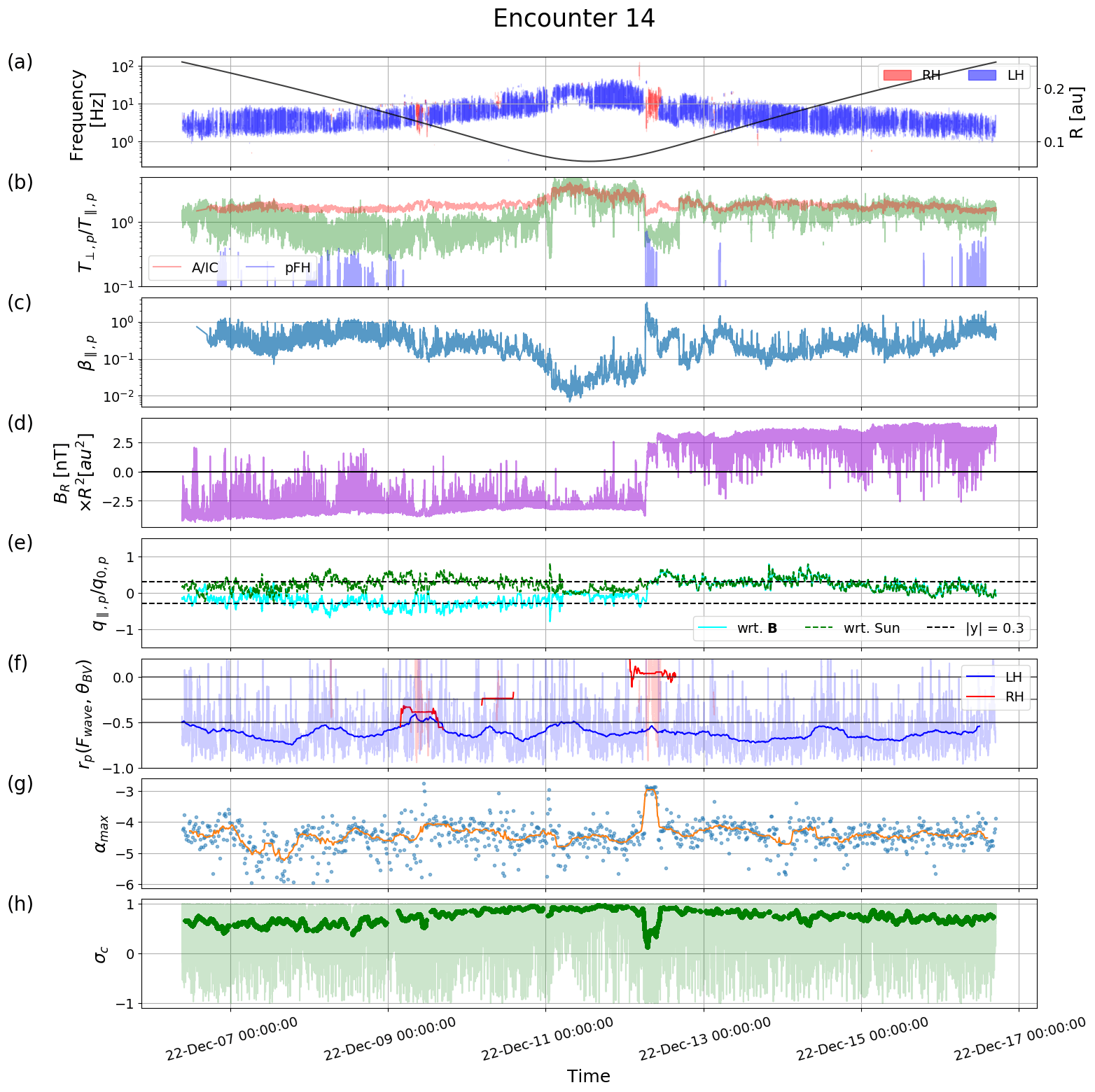}
    \caption{Trends for Encounter 14, organized as Figure \ref{fig:E12_encounterwise_trends}}
    \label{fig:App_Ewise_E14}
\end{figure}

\begin{figure}
    \centering
    \includegraphics[width=0.95\linewidth]{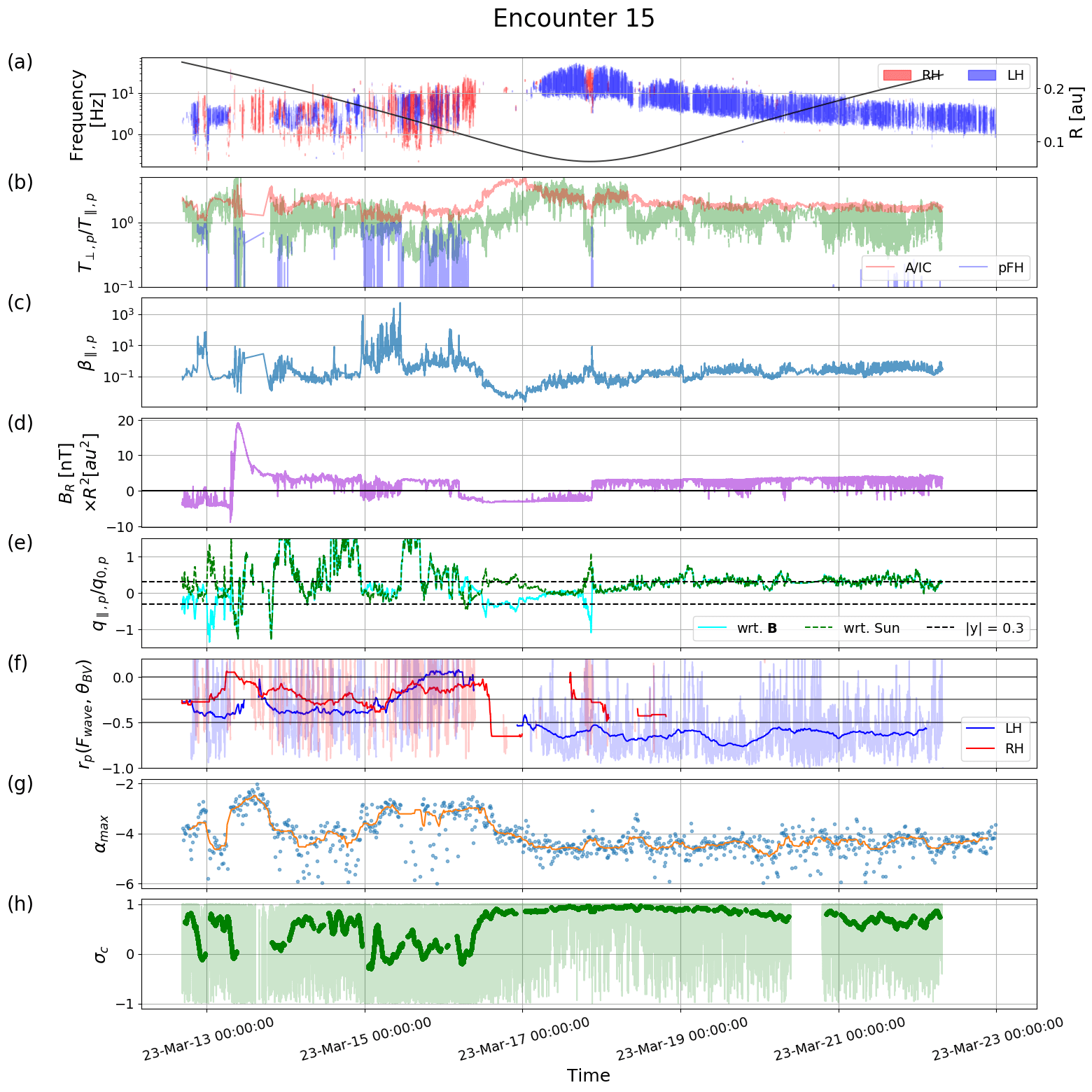}
    \caption{Trends for Encounter 15, organized as Figure \ref{fig:E12_encounterwise_trends}}
    \label{fig:App_Ewise_E15}
\end{figure}

\begin{figure}
    \centering
    \includegraphics[width=0.95\linewidth]{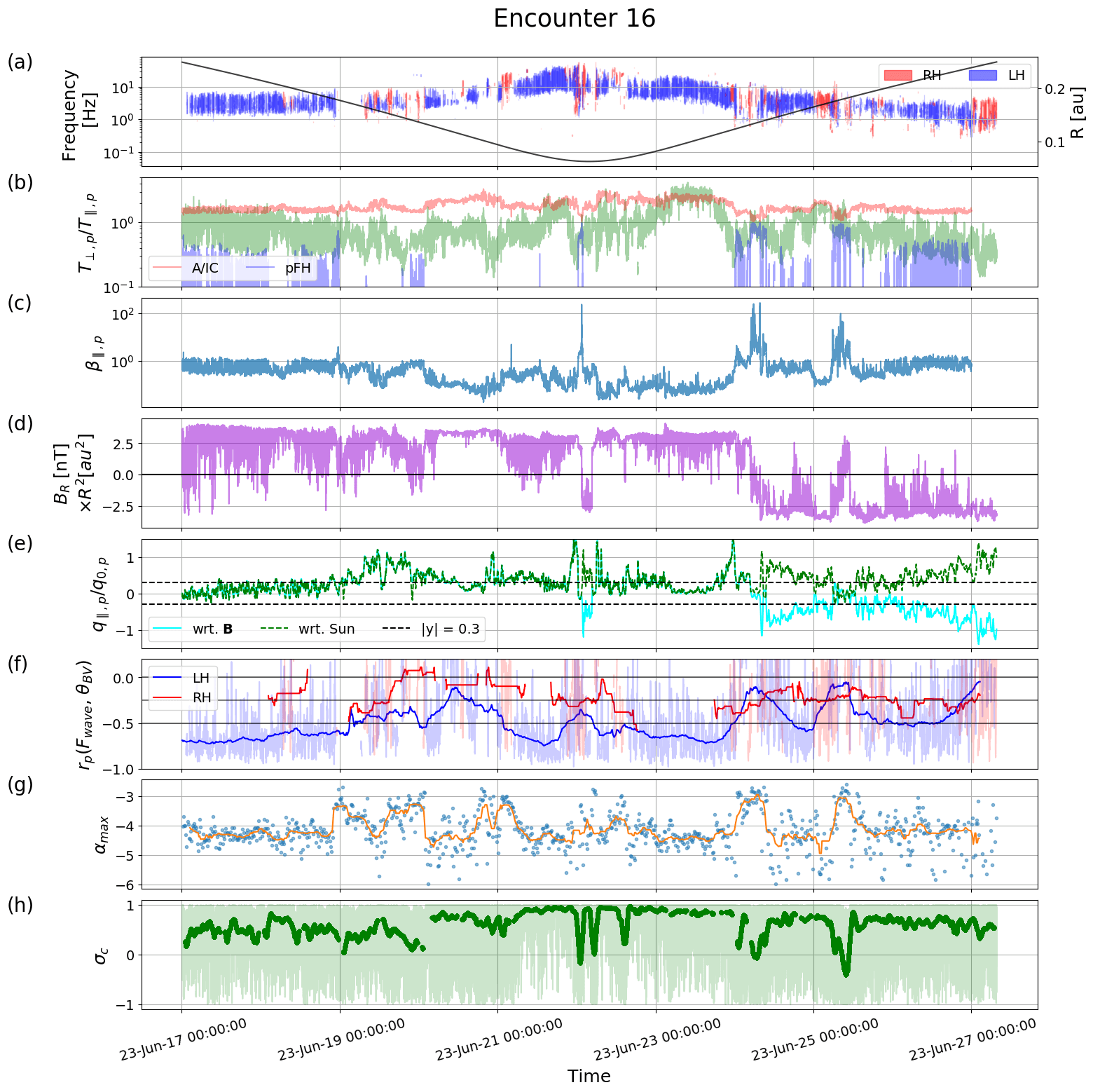}
    \caption{Trends for Encounter 16, organized as Figure \ref{fig:E12_encounterwise_trends}}
    \label{fig:App_Ewise_E16}
\end{figure}

\begin{figure}
    \centering
    \includegraphics[width=0.95\linewidth]{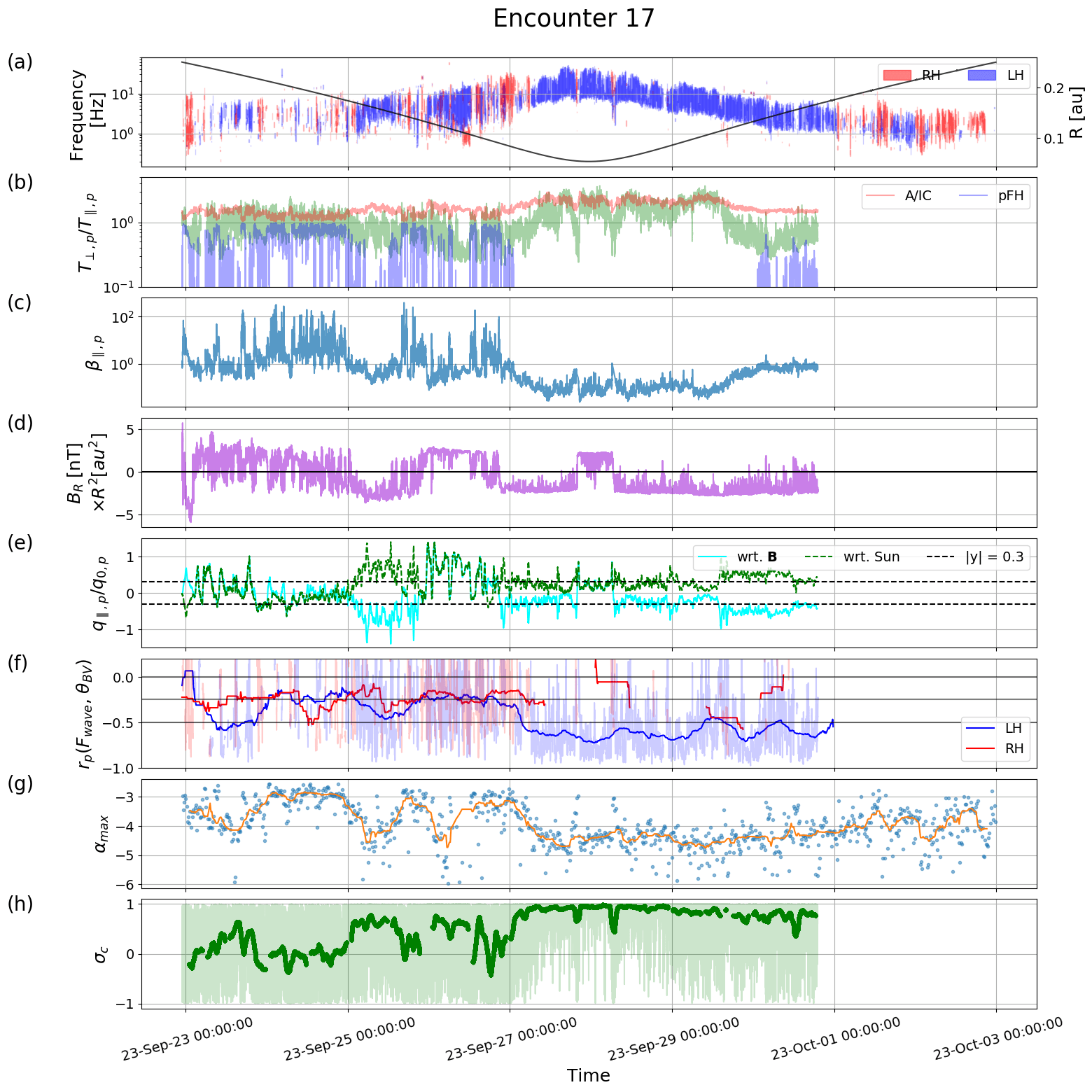}
    \caption{Trends for Encounter 17, organized as Figure \ref{fig:E12_encounterwise_trends}}
    \label{fig:App_Ewise_E17}
\end{figure}

\begin{figure}
    \centering
    \includegraphics[width=0.95\linewidth]{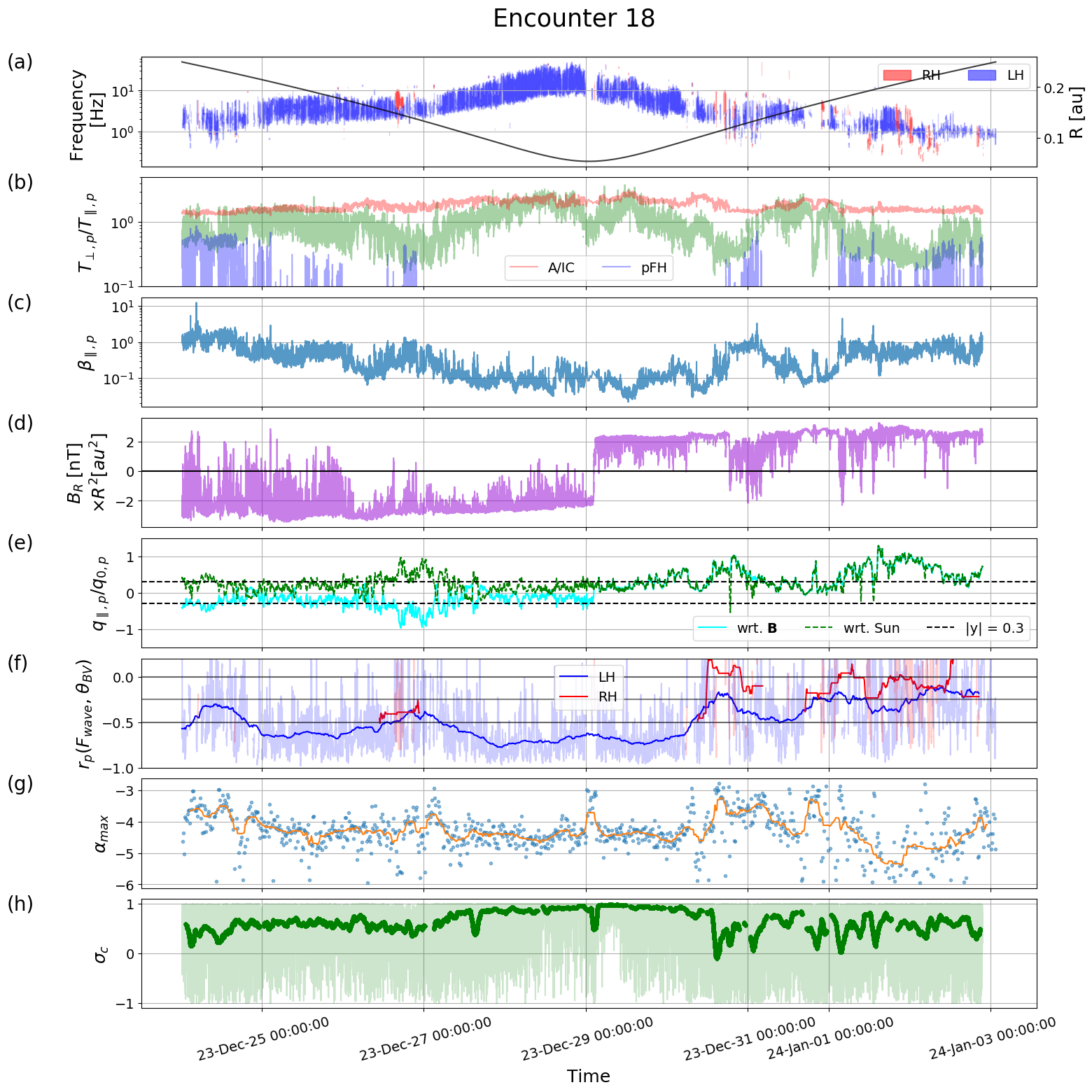}
    \caption{Trends for Encounter 18, organized as Figure \ref{fig:E12_encounterwise_trends}}
    \label{fig:App_Ewise_E18}
\end{figure}

\begin{figure}
    \centering
    \includegraphics[width=0.95\linewidth]{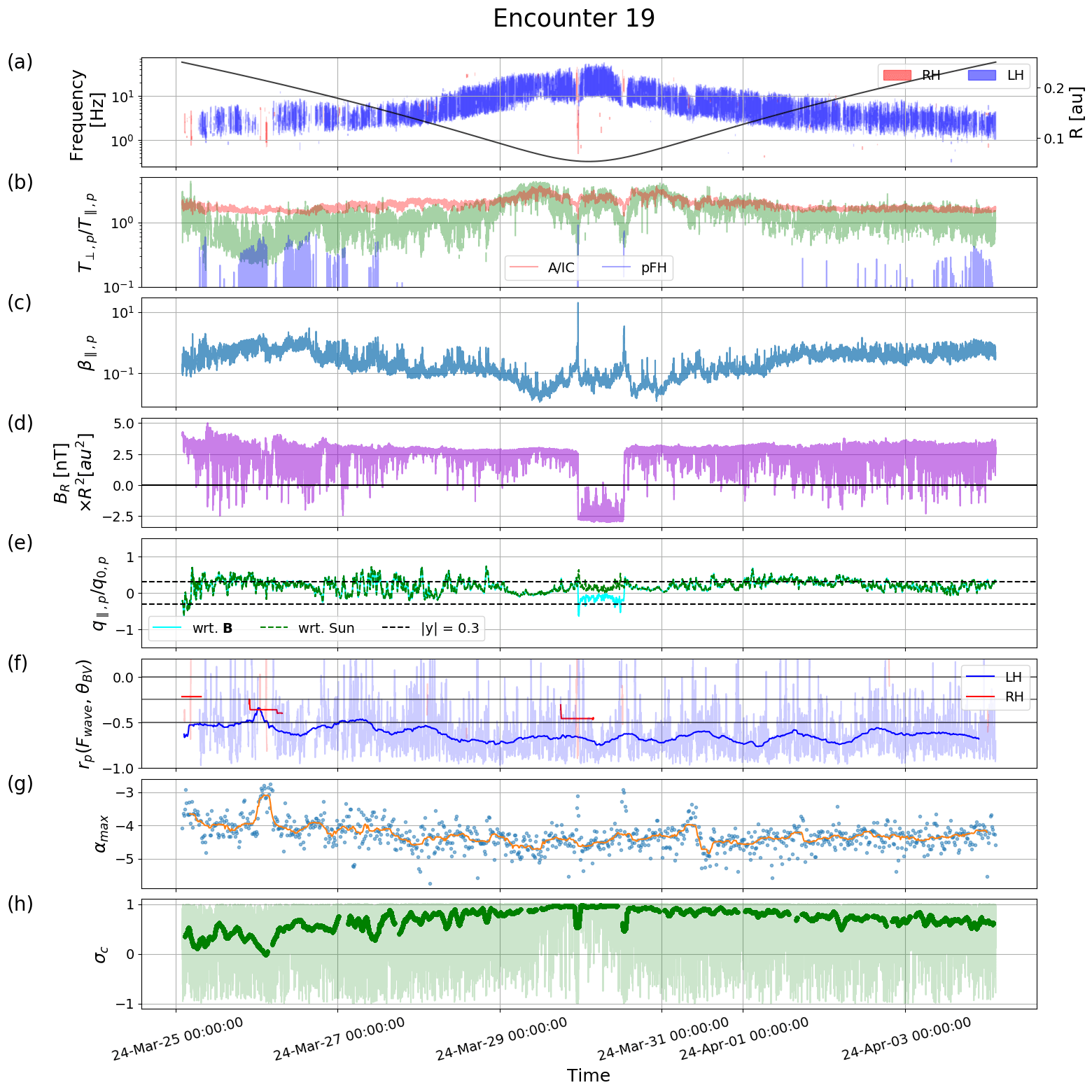}
    \caption{Trends for Encounter 19, organized as Figure \ref{fig:E12_encounterwise_trends}}
    \label{fig:App_Ewise_E19}
\end{figure}

\begin{figure}
    \centering
    \includegraphics[width=0.95\linewidth]{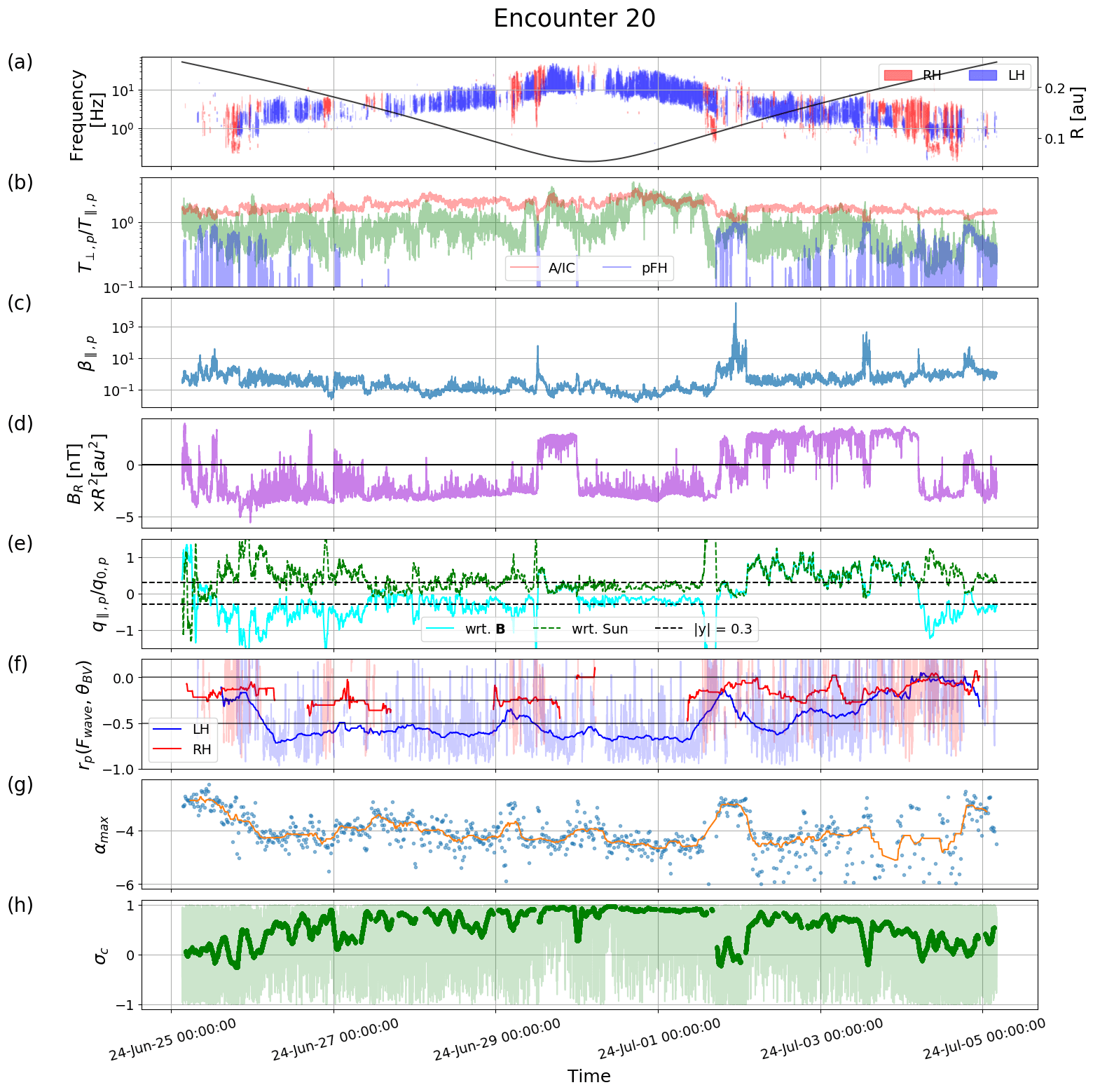}
    \caption{Trends for Encounter 20 organized as Figure \ref{fig:E12_encounterwise_trends}}
    \label{fig:App_Ewise_E20}
\end{figure}

\begin{figure}
    \centering
    \includegraphics[width=0.95\linewidth]{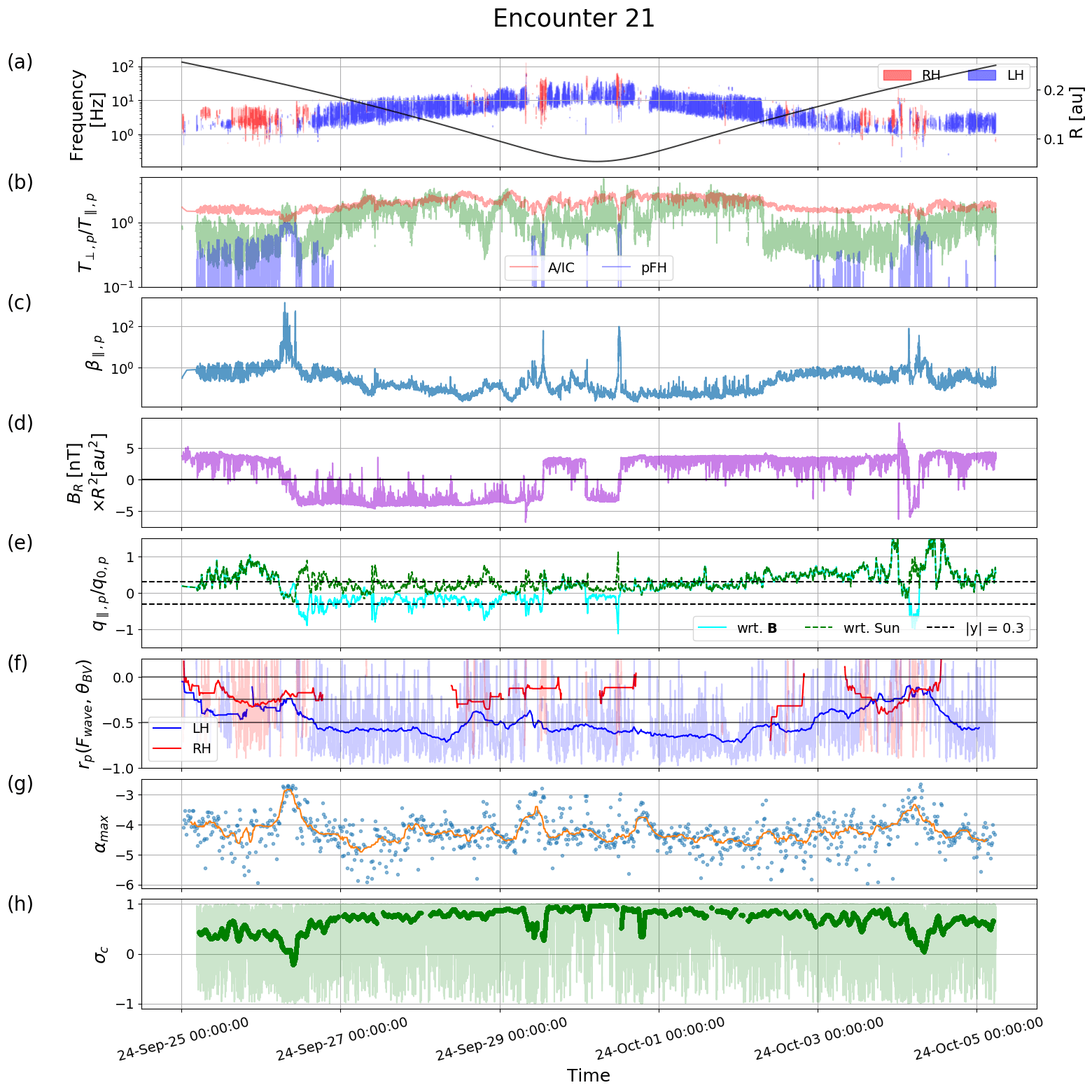}
    \caption{Trends for Encounter 21, organized as Figure \ref{fig:E12_encounterwise_trends}}
    \label{fig:App_Ewise_E21}
\end{figure}

\begin{figure}
    \centering
    \includegraphics[width=0.95\linewidth]{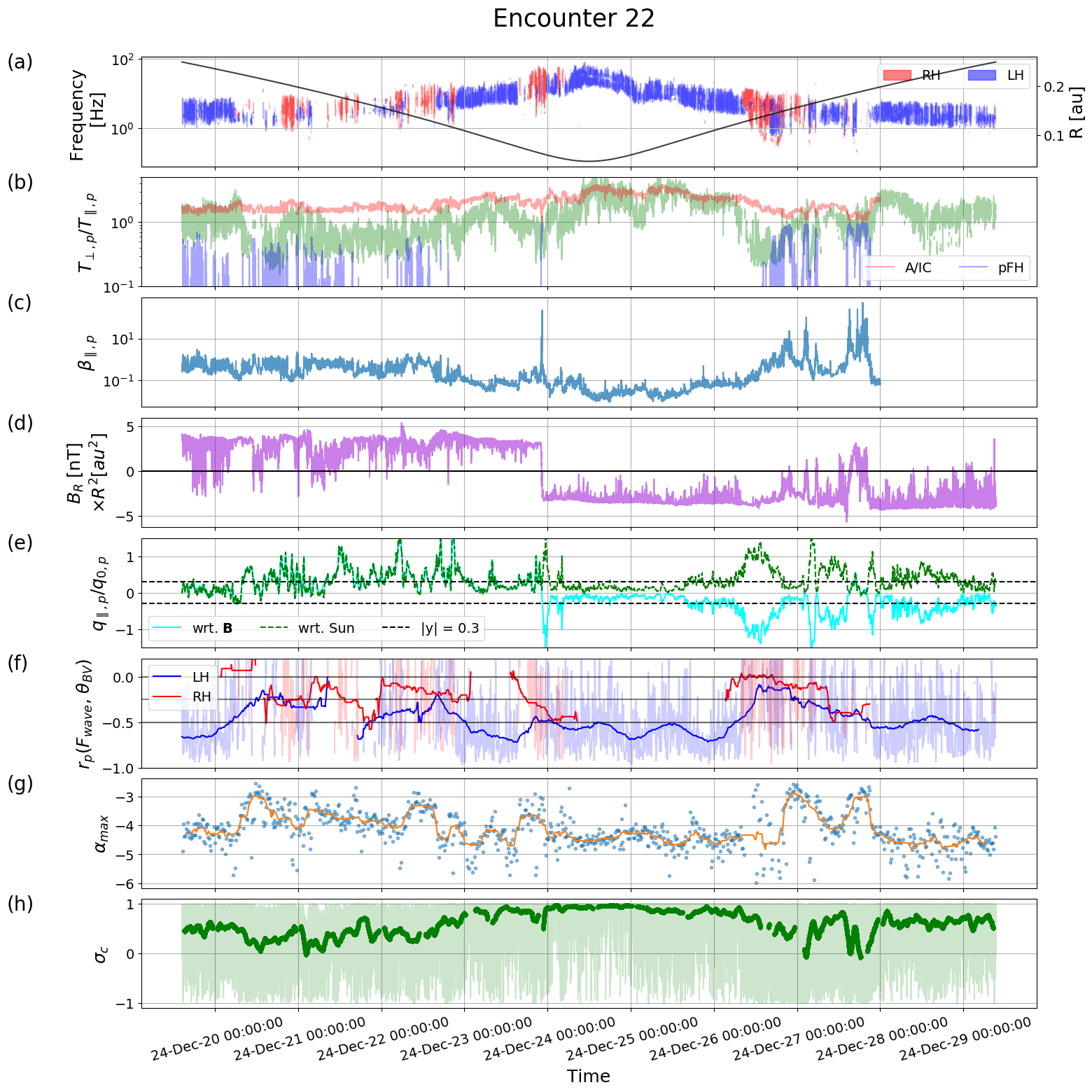}
    \caption{Trends for Encounter 22, organized as Figure \ref{fig:E12_encounterwise_trends}}
    \label{fig:App_Ewise_E22}
\end{figure}

\begin{figure}
    \centering
    \includegraphics[width=0.95\linewidth]{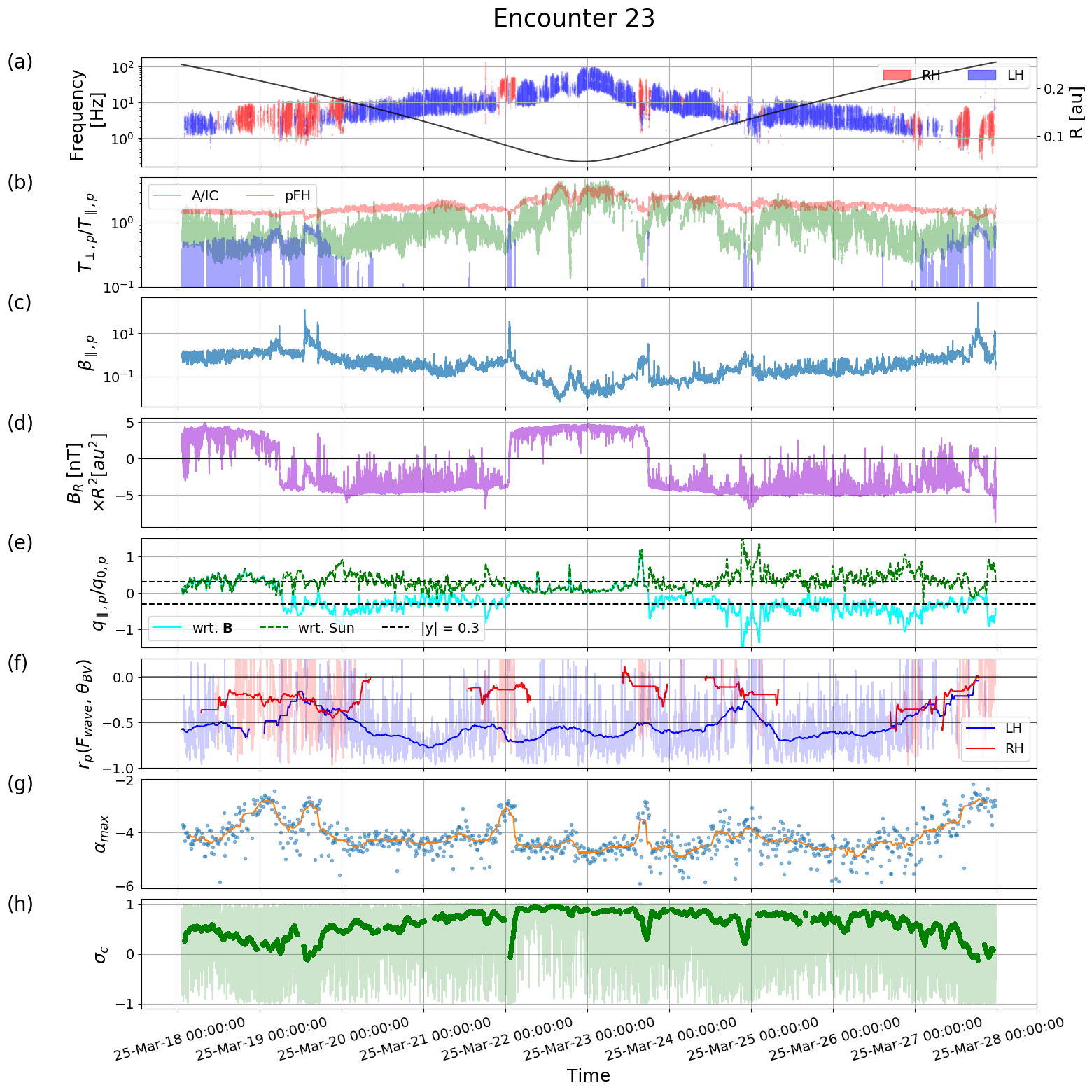}
    \caption{Trends for Encounter 23, organized as Figure \ref{fig:E12_encounterwise_trends}}
    \label{fig:App_Ewise_E23}
\end{figure}

\begin{figure}
    \centering
    \includegraphics[width=0.95\linewidth]{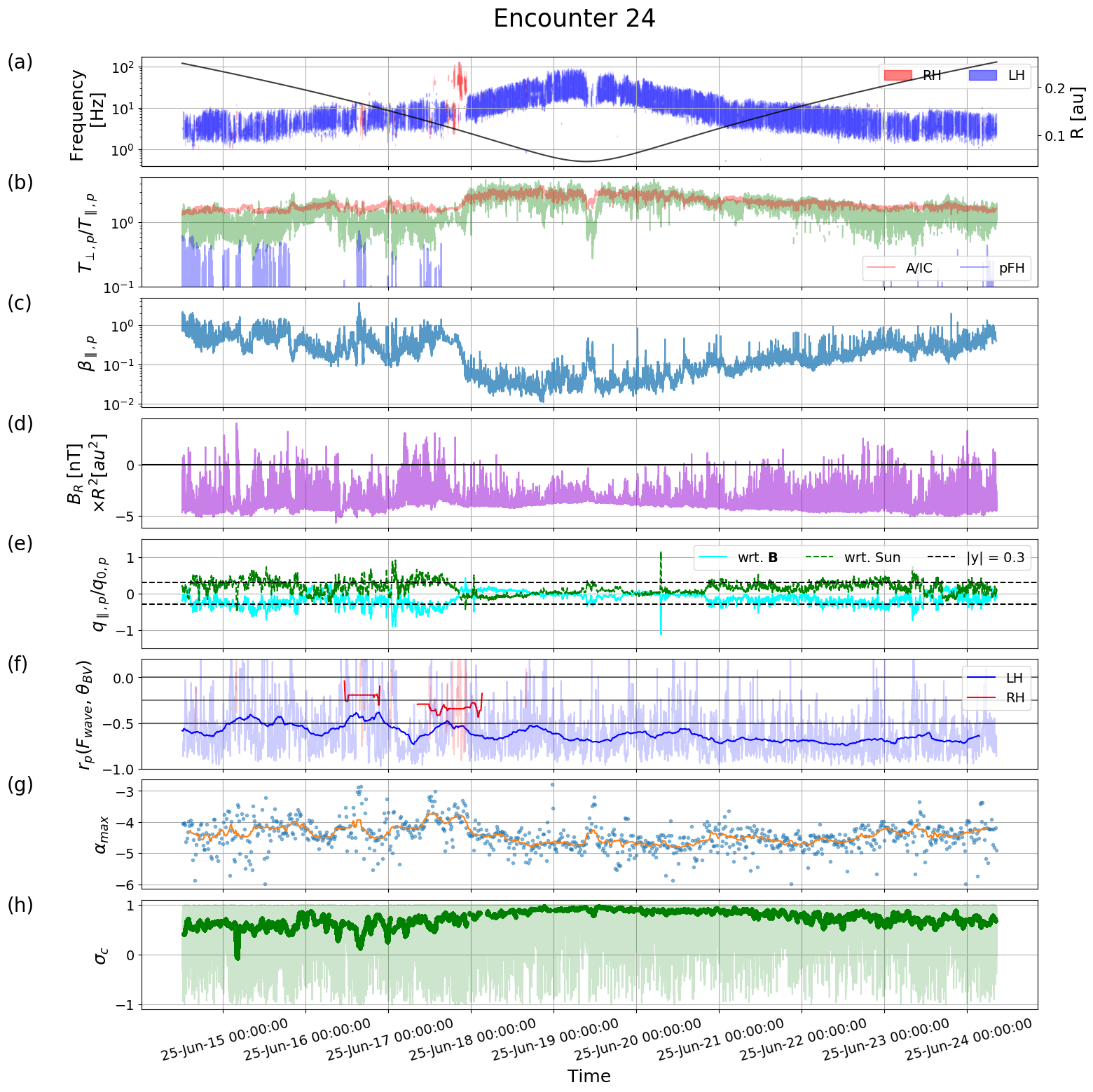}
    \caption{Trends for Encounter 24, organized as Figure \ref{fig:E12_encounterwise_trends}}
    \label{fig:App_Ewise_E24}
\end{figure}

\clearpage
\bibliographystyle{aasjournalv7}
\bibliography{References}{}



\end{document}